\documentclass[3p,12pt]{elsarticle}

\usepackage{scalerel}

\usepackage{amsfonts,amsmath,amssymb,graphicx}
\usepackage{color,mathrsfs,bm,appendix,subfig}
\usepackage[dvipsnames]{xcolor}

\usepackage[figuresright]{rotating}
\usepackage{multirow}
\usepackage{colortbl}

\usepackage[colorlinks=true]{hyperref}

\usepackage{epstopdf}
\usepackage{capt-of}

\usepackage{booktabs}

\DeclareMathOperator*{\A}{\scalerel*{\mathsf{A}}{\sum}}

\journal{Journal of Computational Physics}

\begin{document}

\begin{frontmatter}

\title{A high-order hybridizable discontinuous Galerkin method with fast convergence to steady-state solutions of the gas kinetic equation}

\author{Wei Su, Peng Wang, Yonghao Zhang, Lei Wu\corref{mycorrespondingauthor}}
\address{James Weir Fluids Laboratory, Department of Mechanical and Aerospace Engineering, University of Strathclyde, G1 1XJ Glasgow, United Kingdom}
\cortext[mycorrespondingauthor]{Corresponding author: lei.wu.100@strath.ac.uk}




\begin{abstract}

The mass flow rate of Poiseuille flow of rarefied gas through long ducts of two-dimensional cross-sections with arbitrary shape are critical in the pore-network modeling of gas transport in porous media. In this paper, for the first time, the high-order hybridizable discontinuous Galerkin (HDG) method is used to find the steady-state solution of the linearized Bhatnagar-Gross-Krook equation on two-dimensional triangular meshes. The velocity distribution function and its traces are approximated in the piecewise polynomial space (of degree up to 4) on the triangular meshes and the mesh skeletons, respectively. By employing a numerical flux that is derived from the first-order upwind scheme and imposing its continuity on the mesh skeletons, global systems for unknown traces are obtained with a few coupled degrees of freedom. To achieve fast convergence to the steady-state solution, a diffusion-type equation for flow velocity that is asymptotic-preserving into the fluid dynamic limit is solved by the HDG simultaneously, on the same meshes. The proposed HDG-synthetic iterative scheme is proved to be accurate and efficient. Specifically, for flows in the near-continuum regime, numerical simulations have shown that, to achieve the same level of accuracy, our scheme could be faster than the conventional iterative scheme by two orders of magnitude, while it is faster than the synthetic iterative scheme based on the finite difference discretization in the spatial space by one order of magnitude. The HDG-synthetic iterative scheme is ready to be extended to simulate rarefied gas mixtures and the Boltzmann collision operator.

\end{abstract}


\end{frontmatter}


\section{Introduction}

Accurate physical models and efficient numerical methods are required to describe the gas flow spanning a wide range of gas rarefactions. The conventional Navier-Stokes (NS) equations, however, are valid in the continuum flow regime only, where the Knudsen number $\mathrm{Kn}$, i.e. the ratio of the mean free path of gas molecules $\lambda$ to the flow characteristic dimension $H$, is less than $0.001$. Beyond this regime, gas flows are in strong non-equilibrium and the Boltzmann equation from the gas kinetic theory should be used. According to the Chapman-Enskog expansion, NS equations are the approximated solution of the Boltzmann equation to the first-order of the Knudsen number~\cite{Chapman1970}. As $\mathrm{Kn}$ increases, higher-order terms beyond the linear constitutive relations begin to dominate, and NS equations gradually lose their validity. Not only do the non-equilibrium effects cause velocity slip and temperature jump at solid surface in the slip flow regime ($0.001\le\mathrm{Kn}<0.1$), but also modify the constitutive relations, such as the Newton's law for stress and strain as well as the Fourier's law for heat flux and temperature gradient, in the transition ($0.1\le\mathrm{Kn}<10$) and free-molecular ($10\le\mathrm{Kn}$) flow regimes. In these non-equilibrium flow regimes, the shape of the local velocity distribution function (VDF) of gas molecules is not known a priori. Instead, the VDF should be determined by solving the Boltzmann equation numerically. Two categories of numerical approaches have been developed for this task. One is the direct simulation Monte Carlo method~\cite{Bird1994} that uses a collection of particles to mimic the molecular behavior stochastically, and the other is the deterministic method, which relies on the discretization of the governing equations over computational grids~\cite{Aristov2001}. Generally speaking, the particle-based methods are efficient and robust for high-speed flows, while the deterministic methods are promising for low-speed flows.

In the past decades, due to the rapid development of micro-electro-mechanical systems and the shale gas revolution in North America, extensive works have been devoted to constructing efficient deterministic schemes. These methods often adopt a numerical quadrature to approximate the integration with respect to molecular velocity on a discrete set of velocities~\cite{HUANG1967}. Then, the VDF, which is discrete in the velocity space but continuous in the spatial space and time, is resolved by the finite difference method (FDM), finite volume method (FVM), and finite element method (FEM)~\cite{YANG1995323, KOLOBOV2007589,MIEUSSENS2000429,wu2014}. Compared to the NS equations, numerical simulation of the Boltzmann equation is expensive in terms of computation time and memory consumption. First of all, additional dimensions in the molecular velocity space are discretized,  resulting in a system of governing equations for each discrete velocity. Generally speaking, flows with large values of $\mathrm{Kn}$ require a large number of discrete velocities to resolve the large variations and discontinuities in the VDF~\cite{wu2014,PhysRevE.96.023309}. Second, most of the deterministic schemes treat the streaming and collision separately. Therefore, in order to suppress the numerical diffusion errors, the size of spatial cell and time interval should be smaller than the mean free path and the mean collision time, respectively~\cite{BURT20084653}. For this reason, the deterministic technique becomes costly for near-continuum flows. Finally, the iteration scheme to find steady-state solution converges extremely slowly for flows at low Knudsen numbers, since the exchange of information (e.g. perturbance in the flow field) through streaming becomes very inefficient when binary collisions dominate~\cite{WU2017431}.

Great efforts have been devoted to overcoming the above limitations in various aspects. In addition to the commonly used techniques such as high-order discretization scheme or automatically adaptive refinement in the spatial and velocity spaces~\cite{Arslanbekov2012,BARANGER2014572,ALEKSEENKO2014}, two alternative approaches are worth mentioning here. One is proposed to handle the streaming and collision simultaneously so that the restriction on cell size and time step could be significantly relaxed. This strategy has been realized in the unified gas-kinetic scheme (UGKS)~\cite{XU20107747,LIU2016305,guo2013discrete,guo2015discrete} by calculating the time-evolution of flux at cell interface due to convection and collision. Its advantage of asymptotic-preserving into the NS limit enables UGKS to capture the essential flow physics on coarse grids~\cite{CHEN201552}. Nevertheless, since information is exchanged through the evolution of VDF, UGKS still needs a large number of time steps to obtained steady-state solutions in near-continuum flows~\cite{ZHU201616,WANG201833}. The other strategy, known as the ``synthetic iterative scheme'' (SIS), achieves high efficiency and accuracy in particular with fast convergence property by synchronously solving the kinetic equations and diffusion-type equations for macroscopic quantities~\cite{Valougeorgis2003,Lihnaropoulos2007}. Since the VDF is amended by the macroscopic flow quantities from the diffusion equations at each iterative step, information propagates accurately and fast even on the coarse grid when $\mathrm{Kn}$ is small. Moreover, the macroscopic equations contain high-order moments of VDF to take into account non-equilibrium effects, thus the SIS preserves accuracy in the simulation of high $\mathrm{Kn}$ flows. Based on the FDM in the spatial space, SIS has been successfully applied to Poiseuille flow using the Bhatnagar-Gross-Krook (BGK) kinetic model for single-species gases~\cite{SZALMAS20104315}, and flows of binary and ternary gas mixtures driven by local pressure, temperature and concentration gradients using the McCormack model~\cite{NARIS2004629,Naris2005,SZALMAS201691,SZALMAS201644}. Recently, a SIS is proposed to solve the linearized Boltzmann equation, where the role of realistic intermolecular potentials for gas mixtures in Poiseuille and thermal transpiration flows has been analyzed~\cite{WU2017431}.

In the present paper, to further achieve high-order discretization and enable the capability of dealing with complex geometry, the high-order discontinuous Galerkin (DG) discretization and SIS are coupled to solve the linearized BGK equation for Poiseuille flow through two-dimensional cross-section of arbitrary shape. The developed HDG-SIS has important application in the simulation of rarefied gas flow through complex porous media via the pore-network modeling, where three-dimensional pores with various shapes of two-dimensional cross-sections are extracted~\cite{WRCR:WRCR11219}, e.g. from the ultra-tight shale strata. Accurate and efficient numerical method to solve the gas kinetic equation is urgently needed to find the mass flow rate or apparent permeability of these pores, such that the permeability of the porous media can be obtained by the ``Kirchhoff's circuit law''.

The remainder of the paper is organized as follows. In Sec.~\ref{GKE}, the BGK equation and its synthetic macroscopic equation for the fast convergence of flow velocity in the Poiseuille flow are introduced. In Sec.~\ref{NM} the numerical scheme is described with details in the HDG formulation, flux construction, and implementation of boundary conditions. Four different problems are simulated in Sec.~\ref{Results} to assess the accuracy and efficiency of the proposed HDG-SIS scheme. Conclusions and outlooks are presented in Sec.~\ref{Concludsion}.

\section{The Gas Kinetic Equation}\label{GKE}

The Boltzmann equation describes the evolution of the molecular VDF in dependence of spatial position $\bm x'=(x'_1,x'_2,x'_3)$, molecular velocity $\bm v'=(v'_1,v'_2,v'_3)$, and time $t'$. In Cartesian coordinates it has the form of:
\begin{equation}
\frac{\partial f'}{\partial t'} + \bm v'\cdot\frac{\partial f'}{\partial\bm x'} + \bm a'\cdot\frac{\partial f'}{\partial\bm v'}=\mathcal{C}\left(f'\right).
\end{equation}
Here, $f'$ is the VDF that is defined so that the number density of gas molecules at time $t'$, with velocity lying within the limits $\bm v'$ and $\bm v'+\mathrm{d}\bm v'$, and spatial coordinates lying within $\bm x'$ and $\bm x'+\mathrm{d}\bm x'$, is equal to $f'\mathrm{d}\bm v'\mathrm{d}\bm x'$. $\bm a'=(a'_1,a'_2,a'_3)$ is the external acceleration, while $\mathcal{C}(f')$ is the collision operator, which describes the change in VDF resulting from binary collisions~\cite{Chapman1970}.

Due to complexity of the collision operator, the full Boltzmann equation is amenable to analytical solutions only for few special cases. In practice, deterministic solution is commonly sought for gas kinetic models that reduce $\mathcal{C}(f')$ to simpler collision operators; frequently used are the BGK~\cite{BGK1954}, ellipsoidal statistical BGK~\cite{ESBGK1966}, and Shakhov~\cite{Shakhov1968} models. Here we develop the numerical scheme based on the following BGK equation, which is written in the non-dimensional form as:
\begin{equation}
\frac{\partial f}{\partial t} + \bm v\cdot\frac{\partial f}{\partial\bm x} + \bm a\cdot\frac{\partial f}{\partial\bm v}=\delta\left(F_\text{eq}-f\right),
\end{equation}
where $\bm v$ is $\bm v'$ normalized by the most probable speed $v_\text{m}=\sqrt{2RT_0}$ at the reference temperature $T_0$ with $R$ being the gas constant, $\bm x$ is $\bm x'$ normalized by the characteristic flow length $H$, $\bm a$ is $\bm a'$ normalized by $v_\text{m}^2/H$, $t$ is $t'$ normalized by $H/v_\text{m}$, and $f$ is $f'$ normalized by $n_0/v^3_\text{m}$, where $n_0$ is the average number density of gas molecules at the reference temperature. The normalized equilibrium VDF $F_\text{eq}$ is defined as:
\begin{equation}
F_\text{eq}=\frac{n}{\left(\pi T\right)^{3/2}}\exp\left(-\frac{|\bm v-\bm u|^2}{T}\right),
\end{equation}
where $n$ is the number density of gas molecules normalized by $n_0$, $T$ is the gas temperature normalized by $T_0$, $\bm u=(u_1,u_2,u_3)$ is the macroscopic flow velocity normalized by $v_\text{m}$. Finally, the equivalent rarefaction parameter $\delta$ is defined as the inversed Knudsen number:
\begin{equation}
\delta = \frac{\sqrt{\pi}}{2\mathrm{Kn}}=\frac{p_0H}{\mu_0\sqrt{2RT_0}},
\end{equation}
with $p_0$ and $\mu_0$ being the pressure and shear viscosity of the gas at reference temperature $T_0$, respectively.

When the flow velocity is sufficiently small compared to $v_\text{m}$, and the external acceleration is also small, we can linearize the VDF about the global equilibrium state $f_\text{eq}$ as:
\begin{equation}
f=f_\text{eq}(1+h),\quad f_\text{eq}=\frac{\exp\left(-|\bm v|^2\right)}{\pi^{3/2}},
\end{equation}
and the perturbed VDF $h(\bm x,\bm v)$ is governed by the following linearized BGK equation~\cite{Cercignanibook1988}:
\begin{equation}
\begin{aligned}
\bm v\cdot\frac{\partial h}{\partial\bm x}-2\bm a\cdot\bm v=\mathcal{L}\left(\varrho,\bm u,\tau,\bm v\right)-\delta h,\\
\mathcal{L}\left(\varrho,\bm u,\tau,\bm v\right)=\delta\left[\varrho+2\bm u\cdot\bm v+\tau\left(|\bm v|^2-\frac{3}{2}\right)\right],
\end{aligned}
\label{LBGK}
\end{equation}
in which we have omitted the derivation with respect to the time since we are only interested in the steady-state solution.

The macroscopic gas variables, including the perturbed number density $\varrho$, the flow velocity $\bm u$, and the perturbed temperature $\tau$, are calculated from the velocity moments of the perturbed VDF:
\begin{equation}
\begin{aligned}
\varrho = \int hf_\text{eq}\mathrm{d}\bm v,\quad
\bm u=\int\bm v hf_\text{eq}\mathrm{d}\bm v,\quad
\tau=\frac{2}{3}\int |\bm v|^2hf_\text{eq}\mathrm{d}\bm v-\varrho.
\end{aligned}
\end{equation}

\subsection{Discrete velocity model}


The deterministic approach relies on the discrete velocity method (DVM)~\cite{HUANG1967}, in which a set of $M_\text{v}$ discrete velocities $\bm v^j=(v^{j_1}_1,v^{j_2}_2,v^{j_3}_3)$ are chosen to represent the VDF. If we denote $h^j=h(\bm x,\bm v^j)$, $\mathcal{L}^j\left(\varrho,\bm u,\tau\right)=\mathcal{L}\left(\varrho,\bm u,\tau,\bm v^j\right)$, and $f^j_\text{eq}=f_\text{eq}\left(\bm v^j\right)$, the linearized BGK model equation is replaced by a system of differential equations for $h^j$ that are discrete in the velocity space but still continuous in the spatial space:
\begin{equation}
\bm v^j\cdot\frac{\partial h^j}{\partial\bm x^j}-2\bm a\cdot\bm v^j=\mathcal{L}^j\left(\varrho,\bm u,\tau\right)-\delta h^j,\quad j=1,\dots,M_\text{v}.
\label{DLBGK}
\end{equation}
Then, the macroscopic variables are evaluated using some numerical quadratures:
\begin{equation}
\begin{aligned}
\varrho = \sum^{M_\text{v}}_{j=1}h^jf^j_\text{eq}\omega^j,\quad
\bm u = \sum^{M_\text{v}}_{j=1}\bm v^j h^jf^j_\text{eq}\omega^j,\quad
\tau = \frac{2}{3}\sum^{M_\text{v}}_{j=1}\left(|\bm v^j|^2-\frac{3}{2}\right)h^jf^j_\text{eq}\omega^j,
\end{aligned}
\end{equation}
where $\omega^j$ is the weight of a quadrature rule. Various quadrature rules have been adopted for the selection of discrete velocities and the calculation of VDF moments. Commonly used ones are the Gauss quadrature~\cite{SHIZGAL1981} and the composite Newton-Cote rule with uniform~\cite{YANG1995323} and non-uniform~\cite{wu2014} velocity discretization.

Note that the linearized equilibrium distribution $\mathcal{L}^j$ depends on the macroscopic variables that are evaluated from the unknown perturbed VDF $h^j$. The system of equations (\ref{DLBGK}) are commonly solved by the following implicit iterative scheme:
\begin{equation}
\delta h^{j,(t+1)} + \bm v^j\cdot\frac{\partial h^{j,(t+1)}}{\partial\bm x}=\mathcal{L}^j\left(\varrho^{(t)},\bm u^{(t)},\tau^{(t)}\right)+2\bm a\cdot\bm v^j,
\label{ILBGK}
\end{equation}
where the superscripts $(t)$ and $(t+1)$ represent two consecutive iteration steps. The iteration is terminated when the convergence to the steady solution is achieved. For conciseness, we will omit the index of iteration step in the remainder of the paper unless necessary.


\subsection{The synthetic iterative scheme for asymptotically fast convergence}

It is well known that the iterative scheme~\eqref{ILBGK} is very efficient in the free-molecular flow regime where binary collisions are negligible. However, for near-continuum flows the iteration scheme converges slowly and the results are very likely to be biased by accumulated rounding errors. The accelerated SIS, which has the asymptotic-preserving property in the NS limit and enables rapid convergence to the steady-state, has been developed for the linearized kinetic equations~\cite{Valougeorgis2003,Lihnaropoulos2007,WU2017431} to achieve high efficiency and accuracy.

In this paper, we consider the steady gas flow along a channel of arbitrary cross-section in the $x_1-x_2$ plane, subject to a small pressure gradient in the $x_3$ direction. It is assumed that the channel length is significantly larger than the dimension of its cross-section, thus we can neglect the end effects and consider the flow property depending only on $x_1$ and $x_2$ coordinates. Suppose the pressure gradient are $X_\text{P}$, which is normalized by $p_0/H$, the term $2\bm a\cdot\bm v^j$ in the linearized BGK equation~\eqref{ILBGK} can be replaced by $-X_\text{P}v_3^j$, and the diffusion equation for $u_3$ is given  as~\cite{WU2017431}:
\begin{equation}
\frac{\partial^2 u_3}{\partial x^2_1}+\frac{\partial^2 u_3}{\partial x^2_2}=X_\text{P}\delta -\frac{1}{4}\left(\frac{\partial^2F_{2,0,1}}{\partial x^2_1}+2\frac{\partial^2F_{1,1,1}}{\partial x_1\partial x_2}+\frac{\partial^2F_{0,2,1}}{\partial x^2_2}\right),
\label{Diffusion}
\end{equation}
where
\begin{equation}
F_{m,n,l}(x_1,x_2)=\sum^{M_\text{v}}_{j=1}f^j_\text{eq}h^jH_m(v_1)H_n(v_2)H_l(v_3)\omega^j,
\end{equation}
are high-order moments, with $H_n(v)$ being the $n$-th order physicists' Hermite polynomial.

It should be noted that Eq.~\eqref{Diffusion} is exactly derived from the linearized BGK equation as no approximation is adopted. In the near-continuum flow regime where $\delta$ is large, this equation is reduced to the NS equation. That is to say, it is asymptotic-preserving to the fluid dynamic limit.  Since the diffusion equation exchanges the information very efficiently, fast convergence and high accuracy in the near-continuum flow regime can be easily achieved by solving the gas kinetic equation~\eqref{ILBGK} in parallel with the diffusion equation~\eqref{Diffusion}. On the other hand, when $\delta$ is very small, i.e. the flow is highly rarefied, high-order moments will play significant roles. We assume $X_\text{P}=-1$ in the following calculations.

\section{The HDG Method}\label{NM}

The DG finite element method was initially introduced for the neutron transport equation~\cite{Reed1973}. In the last few decades, after its success in solving nonlinear hyperbolic conservation laws and many convection-dominated problems~\cite{Cockburn1998,Cockburn2001}, this method is recognized as one of the most promising methods for next generation computational fluid dynamics. Similar to the FVM, the DG methods assume discontinuous solution space. The resulting equations are then closed by approximation of the numerical flux on the cell interfaces. Instead of reconstructing the solution on large stencils, high-order spatial accuracy of the DG solution is sought by means of element-by-element polynomial functions. The compactness and their discontinuous nature make the methods ideal for parallelization and the implementation of $hp$-adaptive schemes.

In the recent years, the DG methods have been applied to the gas kinetic model equations~\cite{SU2015123}, and the linearized/full Boltzmann equations~\cite{Gobbert2007,Baker2008,KITZLER20151539} for the simulation of non-equilibrium gas flows. For the kinetic model equations, it has been shown that the second-order DG discretization combined with the explicit Runge-Kutta time iteration is more efficient than the second-order FVM scheme~\cite{SU2015123}. Besides all advantages, the classical DG methods are computationally more expensive than their continuous Galerkin counterparts for steady or implicit schemes. This is largely due to the large number of degrees of freedom in approximating field variables resulting from the discontinuous nature. The shortcoming is enlarged when solving the diffusion equation, where additional auxiliary variables are introduced to approximate the derivatives of the solution~\cite{Cockburn1998local}.

The HDG is then proposed to overcome this disadvantage~\cite{Cockburn2009a}. By producing a final system in terms of the degrees of freedom in approximating traces of the field variables, HDG could significantly reduce the number of global coupled unknowns, since the traces are defined on the cell interfaces and single-valued. Therefore, HDG method is more appropriately used for steady and implicit solvers. This advantage is prominent for the gas kinetic simulation, where a cumbersome system of control equations needed to be resolved. The majority of HDG applications in fluid dynamics to date includes convection-diffusion flow~\cite{Cockburn2009a}, stokes flow~\cite{NGUYEN2010582}, wave propagation problem~\cite{Giorgiani2013} and incompressible/compressible NS flows~\cite{Peraire2010,NGUYEN20111147,Moghtader2016}. Here, for the first time, the HDG method is designed for the gas kinetic equation.

\subsection{Hybridizable discontinuous Galerkin formulation}

We apply the discontinuous Galerkin method to discretize the system in spatial space. Let $\Omega\in\mathbb{R}^2$ be an two-dimensional domain with boundary $\partial\Omega$ in the $x_1-x_2$ plane. Then, $\Omega$ is partitioned in $M_\text{el}$ disjoint regular triangles $\Omega_i$:
\begin{equation}
\Omega=\cup^{M_\text{el}}_i\Omega_i.
\end{equation}
The boundaries $\partial\Omega_i$ of the triangles define a group of $M_\text{fc}$ faces $\Gamma_c$:
\begin{equation}
\Gamma=\cup^{M_\text{el}}_i\{\partial\Omega_i\}=\cup^{M_\text{fc}}_c\{\Gamma_c\}.
\end{equation}

\begin{figure}
	\begin{centering}
		\includegraphics[width=0.75\textwidth]{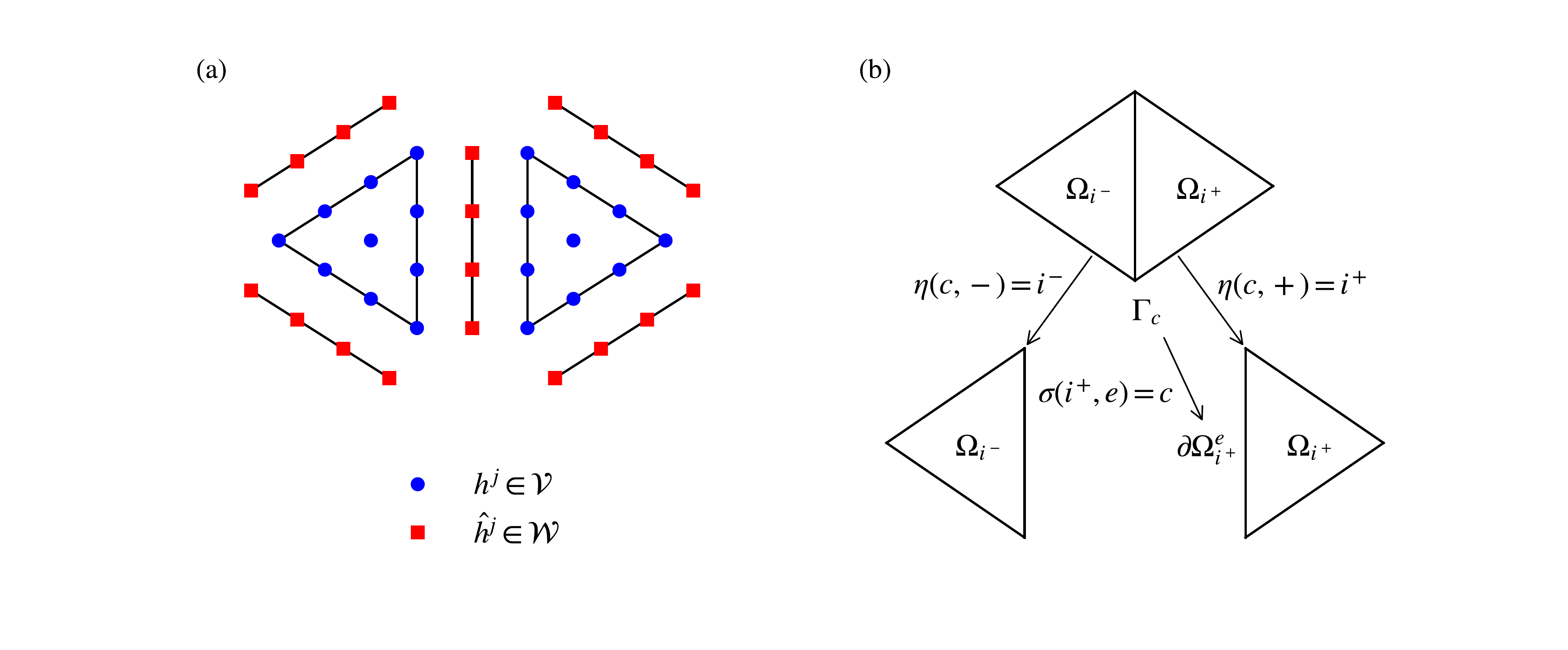}
		\par\end{centering}
	\caption{(a) Nodal points and solution spaces for $k=3$ to approximate $h^j$ and its trace $\hat{h}^j$. (b) Schematic demonstration of the index mapping functions to relate the local edge of a triangle to a global face.}
	\label{Mapping}
\end{figure}

The HDG method provides an approximate solution to $h^j$ on $\Omega_i$ as well as an approximation to its trace $\hat{h}^j$ on $\Gamma_c$ in some piecewise finite element spaces $\mathcal{V}\times\mathcal{W}$ of the following forms:
\begin{equation}
\begin{aligned}[b]
\mathcal{V}=\{\varphi:\ \varphi|_{\Omega_i}\in\mathcal{P}^k(\Omega_i),\ \forall\ \Omega_i\subset\Omega\},\\
\mathcal{W}=\{\psi:\ \psi|_{\Gamma_c}\in\mathcal{P}^k(\Gamma_c),\ \forall\ \Gamma_c\subset\Gamma\},
\end{aligned}
\end{equation}
where $\mathcal{P}^k(D)$ denotes the space of $k-$th order polynomials on a domain $D$, as shown in Fig.~\ref{Mapping}(a). Before describing the HDG formulation, we first define a collection of index mapping functions~\cite{Kirby2012} that allow us to relate the local edge of a triangle, namely $\partial\Omega^e_i$ to a global face $\Gamma_c$. Since the $e$-th edge of the triangle $\partial\Omega_i$ is the $c$-th face $\Gamma_c$, we set $\sigma(i,e)=c$ so that $\partial\Omega^e_i=\Gamma_{\sigma(i,e)}$. Similarly, since the interior face $\Gamma_c\in\Gamma\backslash\partial\Omega$ is the intersection of the two triangles, namely left triangle $\Omega_{i^-}$ and right triangle $\Omega_{i^+}$, we set $\eta(c,+)=i^+$ and $\eta(c,-)=i^-$, then we can denote $\Gamma_c=\partial\Omega_{\eta(c,+)}\cap\partial\Omega_{\eta(c,-)}$. At a boundary face $\Gamma_c\in\partial\Omega$, we say that only the right triangle is involving. The mapping functions are demonstrated in Fig.~\ref{Mapping}(b).

\subsubsection{Formulation of HDG method}

The HDG method solves problem in two steps~\cite{Cockburn2009a}. First, a global problem is setup to determine the trace $\hat{h}^j$ on $\Gamma$. Then, a local problem with $\hat{h}^j$ as boundary condition on $\partial\Omega_i$ is solved element by element to obtain the solutions of $h^j$. Generally speaking, when moving from the interior of the triangle element $\Omega_i$ to its boundary $\partial\Omega_i$, $\hat{h}^j$ defines what the value of $h^j$ on the boundary should be. In the HDG method, it is assumed that $\hat{h}^j$ is singled-valued on each face. 

Introducing  $\left(\cdot\right)$ and $\langle\cdot\rangle$ as $\left(a,b\right)_D=\int_{D\subset\mathbb{R}^2} (a\cdot b)\mathrm{d}x_1\mathrm{d}x_2$ and $\langle a,b\rangle_{D}=\int_{D\subset\mathbb{R}^1}(a\cdot b)\mathrm{d}\Gamma$, respectively, the weak formulation of Eq.~\eqref{ILBGK} for the VDF $h^j$ in each element $\Omega_i$ is:
\begin{equation}
-\left(\nabla\varphi,\bm v^jh^j\right)_{\Omega_i}+\sum^3_{e=1}\langle\varphi,\hat{\bm F}\cdot\bm n\rangle_{\partial\Omega^e_i}+(\varphi,\delta h^j)_{\Omega_i}=(\varphi,s^j)_{\Omega_i}, \quad \text{for all}\ \varphi\in\mathcal{V},
\label{LP}
\end{equation}
where $\hat{\bm F}$ is the numerical trace of the flux,  $\bm n$ is the outward unit normal vector, and $s^j=\mathcal{L}^j-X_\text{P}v_3^{j_3}$. In practice, the numerical trace of the flux is defined as~\cite{Sevilla2016}:
\begin{equation}
\hat{\bm F}^j\cdot\bm n =\bm v^j\cdot\bm n\hat{h}^j+\alpha\left(h^j-\hat{h}^j\right),
\label{Flux}
\end{equation}
where $\alpha$ is a stabilization parameter~\cite{Peraire2010} on each edge $\partial\Omega^e_i$. Here, we evaluate $\alpha$ as:
\begin{equation}
\alpha=|\bm v^j\cdot\bm n|.
\end{equation}

By inserting Eq.~\eqref{Flux} into Eq.\eqref{LP}, we find the solution of $h^j$ on each triangle as a function of the $\hat{h}^j$. In matrix form, it is written as
\begin{equation}
\mathbf{H}^{i,j}=\left[\mathbf{A}^{i,j}\right]^{-1}\mathbf{S}^{i,j}+\left[\mathbf{A}^{i,j}\right]^{-1}\mathbf{\hat{A}}^{i,j}\mathbf{\hat{H}}^{i,j},
\label{SLP}
\end{equation}
where $\mathbf{H}^{i,j}$ ($\mathbf{\hat{H}}^{i,j}$) are the vectors of degrees of freedom of $h^j$ ($\hat{h}^j$) on $\Omega_i$ ($\partial\Omega_i$). The coefficient matrices $\mathbf{A}^{i,j}$, $\mathbf{S}^{i,j}$ and $\mathbf{\hat{A}}^{i,j}$ are given in the Appendix in detail.

The global problem, used for the determination of $\hat{h}^j$, is obtained by imposing the continuity of the normal fluxes at cell interfaces.  For all $\psi\in\mathcal{W}$, the weak formulation is:
\begin{equation}
\begin{aligned}
\langle\psi,\hat{\bm F}\cdot\bm n_{\eta(c,+)}\rangle_{\Gamma_c}+\langle\psi,\hat{\bm F}\cdot\bm n_{\eta(c,-)}\rangle_{\Gamma_c}=0,\quad\text{on}\ \Gamma\backslash\partial\Omega,\\
\langle\psi,\hat{\bm F}\cdot\bm n_{\eta(c,+)}\rangle_{\Gamma_c}+\langle\psi,\hat{\bm G}\cdot\bm n\rangle_{\Gamma_c}=0,\quad\text{on}\ \Gamma\cap\partial\Omega,\\
\end{aligned}
\label{GP}
\end{equation}
where $\hat{\bm F}\cdot\bm n_{\eta(c,\pm)}$ denote the numerical fluxes calculated from the left and right triangles, and $\hat{\bm G}\cdot\bm n$ is the flux defined over the boundary $\partial\Omega$ flowing into the computational domain. Note that the implementation of the boundary condition is equivalent to the standard Neumann boundary condition. By inserting the definition of the numerical flux, i.e. Eq.~\eqref{Flux}, we obtain the matrix system for the global problem:
\begin{equation}
\begin{aligned}
\mathbf{\hat{B}}^{c,j}\mathbf{\hat{H}}^{c,j}=\mathbf{B}^{\eta(c,+),j}\mathbf{H}^{\eta(c,+),j}+\mathbf{B}^{\eta(c,-),j}\mathbf{H}^{\eta(c,-),j},\quad\text{on}\ \Gamma\backslash\partial\Omega,\\
\mathbf{\hat{B}}^{c,j}\mathbf{\hat{H}}^{c,j}=\mathbf{B}^{\eta(c,+),j}\mathbf{H}^{\eta(c,+),j}+\mathbf{\hat{S}}^{c,j},\quad\text{on}\ \Gamma\cap\partial\Omega,
\end{aligned}
\label{GP1}
\end{equation}
where $\mathbf{\hat{H}}^{c,j}$ is the vector of degrees of freedom of $\hat{h}^j$ on $\Gamma_c$. Other coefficient matrices are given in Appendix in detail.

After eliminating the unknowns $\mathbf{H}^{i,j}$ with Eq.~\eqref{SLP} and assembling the Eq.~\eqref{GP1} over all the faces, the global problem becomes:
 \begin{equation}\label{linear}
 \mathbb{K}^j\mathbf{\hat{H}}^j=\mathbb{R}^j,
 \end{equation}
 where $\mathbf{\hat{H}}^j$ is the vector of degrees of freedom of $\hat{h}^j$ on all the faces $\Gamma$, $\mathbb{K}^j$ is the global matrix of the linear system of equations, and $\mathbb{R}^j$ is the vector in the right-hand side of the system.

It is noted that the linear system of equations~\eqref{linear} is highly sparse, in which only face unknowns that involve in two adjacent triangles are coupled at each row. The system could be solved by robust direct solver for sparse unsymmetrical linear systems, e.g. the package PARDISO~\cite{pardiso}. Once the values of $\hat{h}^j$ are obtained, an element-by-element reconstruction of the approximation of $h^j$ is implemented according to Eq.~\eqref{SLP}.

Before describing the implementation of boundary condition, we take an insight into the form of the numerical fluxes. If inserting the expression of flux~\eqref{Flux} into the continuity equation (\ref{GP}) at interior faces, we immediately obtained:
\begin{equation}
\langle\psi,\hat{h}^j\rangle=\frac{1}{2}\langle\psi, h^j_{\eta(c,+)}+h^j_{\eta(c,-)}\rangle.
\end{equation}
That is, the trace $\hat{h}^j$ at interior face is equal, in a weak sense, to the average of $h^j_{\eta(c,\pm)}$, which are evaluated at the interface from the left and right triangles, respectively. Then we obtain an equivalent expression for $\hat{\bm F}\cdot\bm n$:
\begin{equation}
\hat{\bm F}\cdot\bm n_{\eta(c,\pm)}=\begin{cases}
\bm v^j\cdot\bm n_{\eta(c,\pm)}h^j_{\eta(c,\pm)},\quad\bm v^j\cdot\bm n_{\eta(c,\pm)}\ge0\\
\bm v^j\cdot\bm n_{\eta(c,\pm)}h^j_{\eta(c,\mp)},\quad\bm v^j\cdot\bm n_{\eta(c,\pm)}<0
\end{cases},
\end{equation}
which is exactly the upwind scheme.

\subsubsection{Implementation of boundary condition}
In order to complete the formulation, we need to specify the flux $\hat{\bm G}\cdot\bm n$ at boundary $\partial\Omega$. To be consistent with the evaluation the fluxes at interior faces, we calculate the boundary flux as:
\begin{equation}
\hat{\bm G}\cdot\bm n = \bm v^j\cdot\bm n\hat{h}^j+\alpha\left(g^j-\hat{h}^j\right),
\end{equation}
where $g^j$ is the boundary value of $h^j$ and $\bm n$ is the outward unit normal vector at the boundary pointing into the flow field. In this paper, the fully diffuse boundary condition is used to determine the perturbed VDF $g^j$ at the solid surface. Suppose the solid wall is static and has the temperature $T_0$, the perturbed VDF for the reflected molecules at the wall (i.e., when $\bm v^j\cdot\bm n>0$) is given by
$g^j=-2\sqrt{\pi}\sum_{\bm v^j\cdot\bm n<0}\bm (\bm v^j\cdot\bm n)f^j_\text{eq}h^j\omega^j$, which is always zero in this specific problem where $h(v_3)=-h(-v_3)$.

Other type of boundary conditions, such as the Maxwell diffuse-specular boundary condition with given tangential momentum accommodation coefficient, symmetry boundary, periodic boundaries, as well as far-pressure inlet/outlet boundary could be incorporated straightforwardly~\cite{SU2015123,wu_struchtrup_2017}.

\subsection{HDG for the synthetic equation}\label{SIS}

The HDG method for solving the diffusion equation has been well developed~\cite{Cockburn2009a,Sevilla2016}, in which two auxilliary variables are introduced to approximate the derivatives of $u_3$, thus HDG approximation is synchronously taken for the flow velocity $u_3$, its derivatives $\nabla u_3$, and its trace $\hat{u}_3$. Here, we skip the details of the scheme, and discuss several modifications that we tailored for the current problem.

First of all, since the second-order partial derivatives of the high-order moments also appear in the equation~\eqref{Diffusion}, we rewrite the equation into a first-order system in the form as:
\begin{equation}
\begin{aligned}
\nabla\cdot\bm q=X_\text{P}\delta,\\
\bm q+\nabla u_3+\bm r=\bm 0,
\end{aligned}
\label{DE}
\end{equation}
where the vector $\bm{r}$ is
\begin{equation}
\bm r=\frac{1}{4}\left[\frac{\partial F_{2,0,1}}{\partial x_1}+\frac{\partial F_{1,1,1}}{\partial x_2},\frac{\partial F_{1,1,1}}{\partial x_1}+\frac{\partial F_{0,2,1}}{\partial x_2}\right]^T.
\end{equation}
That is, the introduced auxilliary variable $\bm q$ is used to approximate the combination of the derivatives of $u_3$ and high-order moments, which  guarantees the stability and local solvability of the auxiliary variables.

Second, to specify the boundary condition of $\hat{u}_3$, we evaluate it from the perturbed VDF as:
\begin{equation}
\langle\psi,\hat{u}_3\rangle_{\Gamma_c}=\langle\psi,\sum^N_jv^{j_3}_3f^j_\text{eq}h^j\omega^j\rangle_{\Gamma_c},\quad\text{on}\ \Gamma\cap\partial\Omega.
\label{BCU}
\end{equation}
This could guarantee the proper value of the flow velocity at boundary, especially when the slip velocity at the solid surface is large for highly rarefied flow.

We state the procedures of SIS for the linearized BGK equation as follows:
\begin{itemize}
\item {When $h^{j,(t)}$ and $u^{(t)}_3$ are known at the $t$-th iteration step, calculate the VDF $h^{j,(t+1)}$ at $(t+1)$-th step by solving Eq. (\ref{ILBGK});}
\item{From $h^{j,(t+1)}$, calculate the high-order moments $F_{2,0,1}$, $F_{1,1,1}$ and $F_{0,2,1}$;}
\item{From $h^{j,(t+1)}$, calculate the flow velocity trace $\hat{u}^{(t+1)}_3$ at boundary, see Eq. (\ref{BCU});}
\item{Calculate $u^{(t+1)}_3$ by solving the diffusion equation (\ref{DE}), with the boundary condition obtained from the previous step.}
\end{itemize}
The above iterative procedure is continued until the steady-state is reached. For the following calculation, the stabilization parameter appears in the expression of numerical flux for $\hat{\bm q}\cdot\bm n$ (Eq. (8) in reference~\cite{Sevilla2016}) is set to be 1.

\section{Results and Discussions}\label{Results}

The HDG method of $k$ up to 4 is applied to solve the linearized BGK kinetic model equation~\eqref{ILBGK} in parallel with the diffusion equation~\eqref{Diffusion}. The convergence criterion for the iterative procedure described in Sec.~\ref{SIS} is that the global relative residual in flow velocity between two successive iteration steps is less than $10^{-5}$. The residual is defined as
\begin{equation}
R=\frac{|\int u^{(t+1)}_3-u^{(t)}_3\mathrm{d}x_1\mathrm{d}x_2|}{|\int u^{(t)}_3\mathrm{d}x_1\mathrm{d}x_2|}.
\end{equation}

\begin{figure}
	\begin{centering}
		\includegraphics[width=0.9\textwidth]{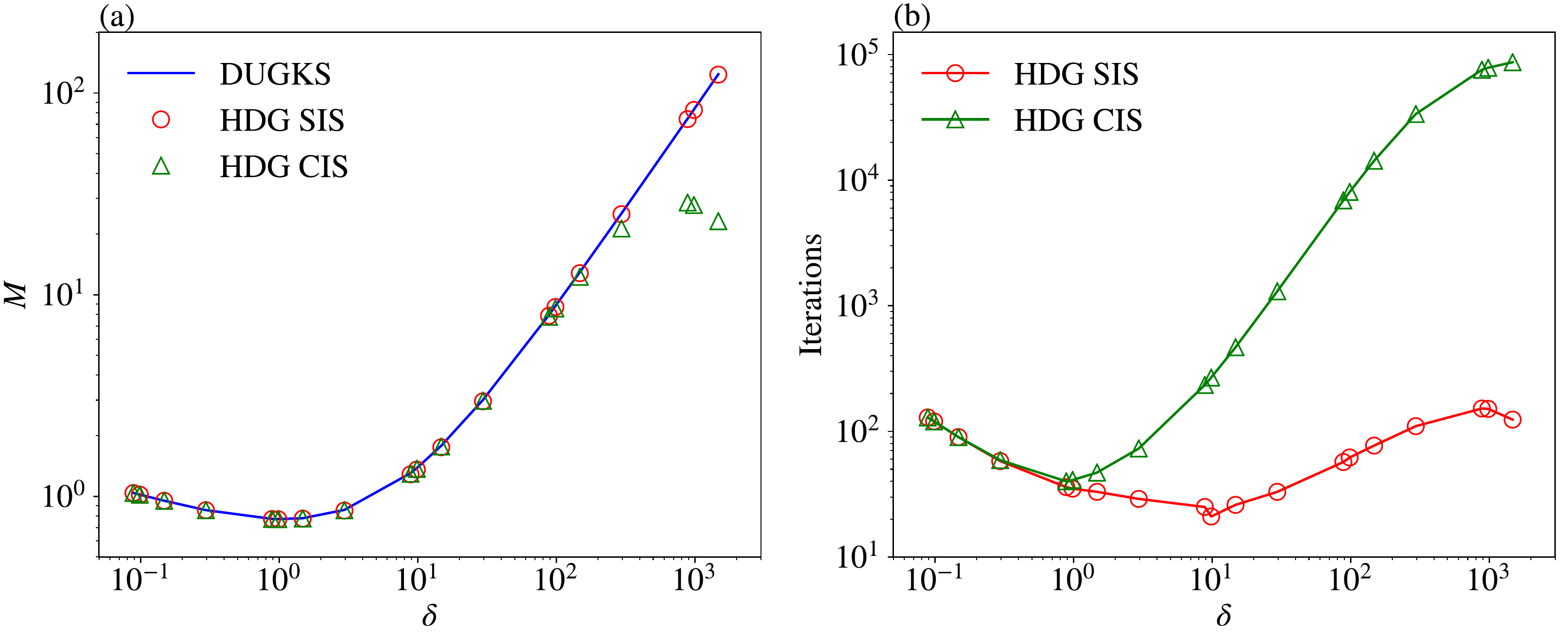}
		\par\end{centering}
	\caption{Comparison of the CIS and SIS: Poiseuille flow between two parallel plates over a wide range of rarefaction. (a) Mass flow rates; (b) Number of iteration steps.}
	\label{Poi_MFR}
\end{figure}

In addition to the profiles of flow velocity, we are interested in the property of dimensionless mass flow rate (MFR):
\begin{equation}
M=\frac{1}{H^2}\int\int u_3\mathrm{d}x_1\mathrm{d}x_2.
\end{equation}

To assess the accuracy and efficiency of the proposed scheme, our numerical results are compared with the discrete UGKS (DUGKS) solutions, which have been verified in all the flow regimes~\cite{WANG201833}, or available data from literature. In the four test cases below, the convergence tests in terms of the discrete velocities are performed first to determine the number of points in the molecular velocity space: the convergence is said to be reached if further refinement of the velocity grid would only improve the solutions by a magnitude no more than 0.5\%.  The entire tests are done in double precision on a workstation with Intel Xeon-E5-2680 processors and 132 GB RAM. During iteration, we call the relative routines in Intel Math Kernel Library (MKL) to invert the matrix. Moreover, to solve the HDG global equations, we call the direct sparse solver, Intel MKL PARDISO.

\subsection{Fast convergence of the SIS: Poiseuille flow between two parallel plates}

The Poiseuille flow between two parallel plates with a distance of $H$ is used to assess the accuracy and fast convergence of the proposed HDG solver. The one-dimensional flow is resolved on a two-dimensional domain of $\Omega=[0,0.5]\times[0,1.0]$ with 4 uniform isosceles right triangles being set along the direction perpendicular to the plates, say, the $x_2$ direction. Therefore, the height of each triangle is equal to 0.354, which is larger than the mean free path when $\mathrm{Kn}<0.354$ or equivalently $\delta>2.50$.

The MFR at different rarefaction parameter $\delta$, obtained from the SIS with $k=3$, is illustrated in Fig.~\ref{Poi_MFR}(a) and compared with those from the DUGKS and the conventional iteration scheme (CIS). In the CIS, only the linearized BGK model equation~\eqref{ILBGK} is solved. The calculation parameters including the numbers of grid points employed in the DUGKS could be found in the relevant reference~\cite{WANG201833}. It is shown that, the MFR first drops to the minimum value at $\delta\sim1$ and then rapidly increases with $\delta$. The Knudsen minimum of $M$ is due to the competition of two effects: when degree of rarefaction increases, the slip velocity at the plates becomes larger, while the velocity profile becomes flatter~\cite{PhysRevE.96.023309}. The SIS could obtain MFRs with high accuracy on such a coarse grid over a wide range of flow regimes. The relative $L_2$ errors of the SIS results to the ones of the DUGKS are within 1.1\%. However, the CIS results possess obvious errors when $\delta\gtrsim150$. For example, the MFR from the CIS is about $61.7\%$ smaller than that of the DUGKS at $\delta=886.2$. This is due to the fact that the spatial resolution is too low such that the numerical viscosity is not negligible in comparison with the physical viscosity of the gas in the CIS, while in the SIS the macroscopic diffusion equation~\eqref{Diffusion} is solved with the physical viscosity.

\begin{table}
\caption{Comparisons between the CIS and SIS in terms of the accuracy, the number of iterations (Itr denotes the number of iteration steps to reach the convergence criterion $R<10^{-5}$), and the CPU time $t_\text{c}$. The Poiseuille flow between two parallel plates is considered. }

\centering{}%
\begin{tabular}{ccccccccc}
\hline
\multirow{2}{*}{$\delta$} & \multirow{2}{*}{$k$} & \multicolumn{3}{c}{CIS} &  & \multicolumn{3}{c}{SIS}\tabularnewline
\cline{3-5} \cline{7-9}
 &  & $L_{2}$ error & Itr & $t_\text{c}$, {[}s{]} &  & $L_{2}$ error & Itr & $t_\text{c}$, {[}s{]}\tabularnewline
\hline
\multirow{4}{*}{$88.62$} & 1 & $2.16\times10^{-1}$ & 6121 & 7532.1 &  & $3.91\times10^{0}$ & 210 & 264.7\tabularnewline
 & 2 & $2.29\times10^{-2}$ & 6886 & 11541.5 &  & $1.05\times10^{-2}$ & 85 & 152.8\tabularnewline
 & 3 & $2.28\times10^{-2}$ & 6896 & 14855.3 &  & $1.05\times10^{-2}$ & 57 & 134.9\tabularnewline
 & 4 & $2.27\times10^{-2}$ & 6896 & 22729.5 &  & $1.01\times10^{-2}$ & 44 & 158.7\tabularnewline
\hline
\multirow{4}{*}{8.862} & 1 & $6.78\times10^{-2}$ & 224 & 554.5 &  & $3.17\times10^{-1}$ & 45 & 99.8\tabularnewline
 & 2 & $7.80\times10^{-3}$ & 234 & 810.3 &  & $2.10\times10^{-2}$ & 30 & 92.2\tabularnewline
 & 3 & $7.21\times10^{-3}$ & 234 & 1068.8 &  & $1.35\times10^{-2}$ & 25 & 100.0\tabularnewline
 & 4 & $7.01\times10^{-3}$ & 234 & 1553.2 &  & $1.01\times10^{-2}$ & 23 & 149.9\tabularnewline
\hline
\multirow{4}{*}{0.8862} & 1 & $7.65\times10^{-3}$ & 40 & 104.5 &  & $1.90\times10^{-3}$ & 36 & 83.3\tabularnewline
 & 2 & $2.04\times10^{-3}$ & 40 & 160.2 &  & $4.21\times10^{-3}$ & 36 & 116.2\tabularnewline
 & 3 & $2.00\times10^{-3}$ & 40 & 201.1 &  & $2.51\times10^{-3}$ & 36 & 155.7\tabularnewline
 & 4 & $1.99\times10^{-3}$ & 40 & 282.9 &  & $2.17\times10^{-3}$ & 36 & 236.6\tabularnewline
\hline
\multirow{4}{*}{0.08862} & 1 & $1.92\times10^{-3}$ & 129 & 328.6 &  & $2.12\times10^{-3}$ & 129 & 322.4\tabularnewline
 & 2 & $8.94\times10^{-4}$ & 130 & 454.2 &  & $9.37\times10^{-4}$ & 129 & 454.3\tabularnewline
 & 3 & $9.14\times10^{-4}$ & 129 & 562.9 &  & $9.20\times10^{-4}$ & 129 & 605.5\tabularnewline
 & 4 & $9.13\times10^{-4}$ & 129 & 828.9 &  & $9.14\times10^{-4}$ & 129 & 936.1\tabularnewline
\hline
\end{tabular}
\label{Poi}
\end{table}

Another superiority of the SIS to the CIS is immediately seen from Fig.~\ref{Poi_MFR}(b), which shows the iteration steps to reach the steady-state solution for both CIS and SIS. When the CIS is used, the number of iteration steps increases rapidly with the rarefaction parameter in the near-continuum flow regime ($\delta\ge10$), whereas those of the SIS only increases slightly. In the late transition flow regime ($\delta<1$), however, the number of iterative steps are almost the same for both schemes. This is further confirmed in Table~\ref{Poi}, where the relative $L_2$ error of MFRs (calculated based on the DUGKS results), the number of iteration steps, and the total CPU time are listed for various rarefaction parameters $\delta$ and degrees of approximation polynomials in the HDG method. For each case at $\delta=88.62$, $20$ uniform points were used to discretize the velocity space truncated in the range of $[-4,4]$ in each direction, while $24$ non-uniform points~\cite{PhysRevE.96.023309} were employed for other cases. It is interesting to note that with the same number of triangles, the number of iterative steps of the CIS reaches a constant value as the degree of polynomials in the HDG discretization increases. While at large $\delta$, the number of iterative steps of the SIS further drops as higher degree of approximation polynomials is employed. Compared to the kinetic equation the time to solve Eq.~\eqref{Diffusion} is negligible, the CPU time saving is proportional to the the reduction of iteration steps. Therefore, the SIS needs significantly less time to reach converged solutions than the CIS. At $\delta=8.862$, the SIS with $k=4$ is 10 times faster than the CIS, while at $\delta=88.62$ it is 143 times faster.

\begin{figure}
	\centering
		\includegraphics[width=0.9\textwidth]{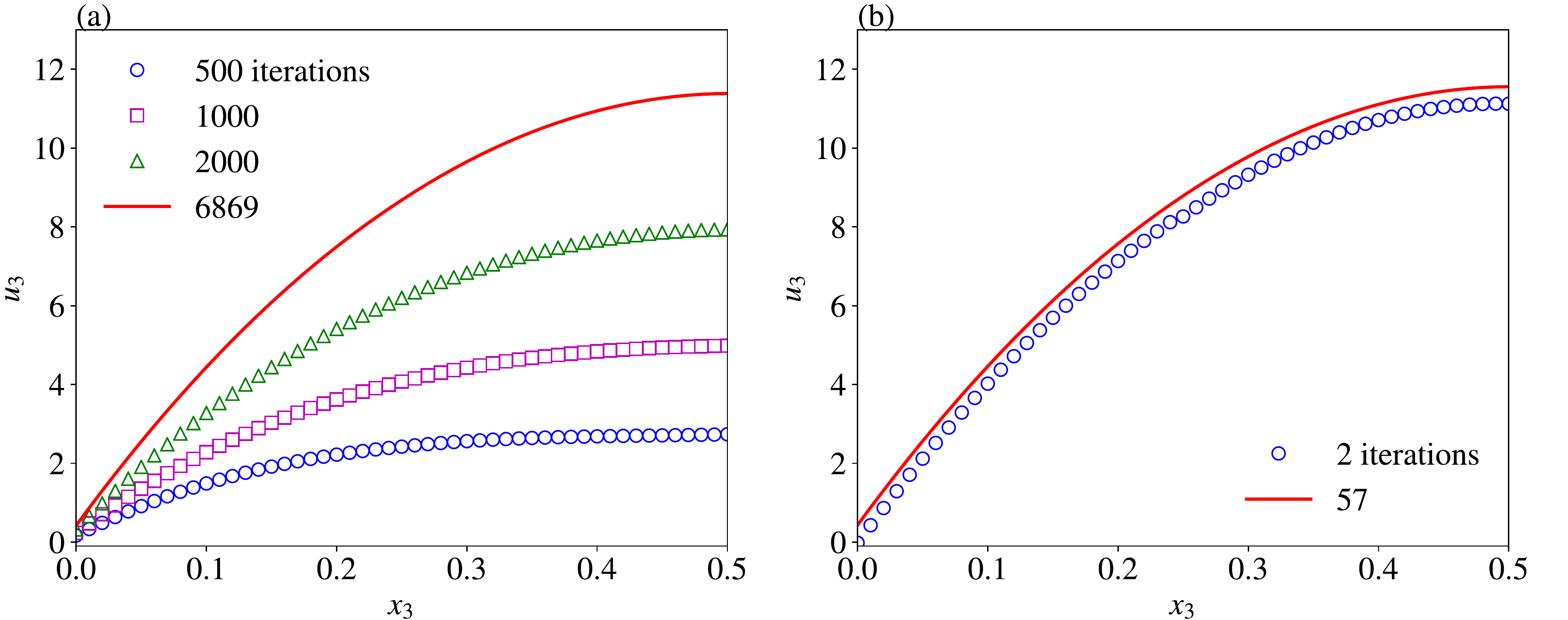}
	\caption{Comparison of the CIS (a) and SIS (b) in terms of the convergence history for the velocity in Poiseuille flow between two parallel plates. The rarefaction parameter is $\delta=88.62$, and the order of HDG is $k=3$. Red lines are the converged result.}
	\label{Poi_U}
\end{figure}

To show how the SIS works in the near-continuum flow regime, the convergence histories of the SIS and CIS are plotted in Fig.~\ref{Poi_U} when $\delta=88.62$. Staring from the zero disturbance, the flow velocity gradually increases from zero due to the gas-gas and gas-surface collisions. From Fig.~\ref{Poi_U}(a) we see that, near the wall the flow velocity quickly approaches the converged value, while the velocity in the bulk adjusts rather slowly. That is to say, due to the frequent molecular collisions, the external acceleration from the imposed pressure gradient slowly penetrate the bulk flow filed. As a result, a large number of iterations is required in the CIS to promote the flow velocity reaching to the maximum value. However, this situation is changed in the SIS, where the macroscopic diffusion equation~\eqref{Diffusion} quickly generate the parabolic velocity profile (the second-order derivative $\partial^2u_3/\partial x_2^2$ is very close to $-\delta$) in the bulk, which boots the convergence significantly. From Fig.~\ref{Poi_U}(b) it is found that the velocity profile of the SIS is already very closed to the final solution, even at the second iterative step.

\subsection{Comparison of the HDG and FDM: flow along a channel of square cross-section}

\begin{figure}
	\begin{centering}
		\includegraphics[width=0.95\textwidth]{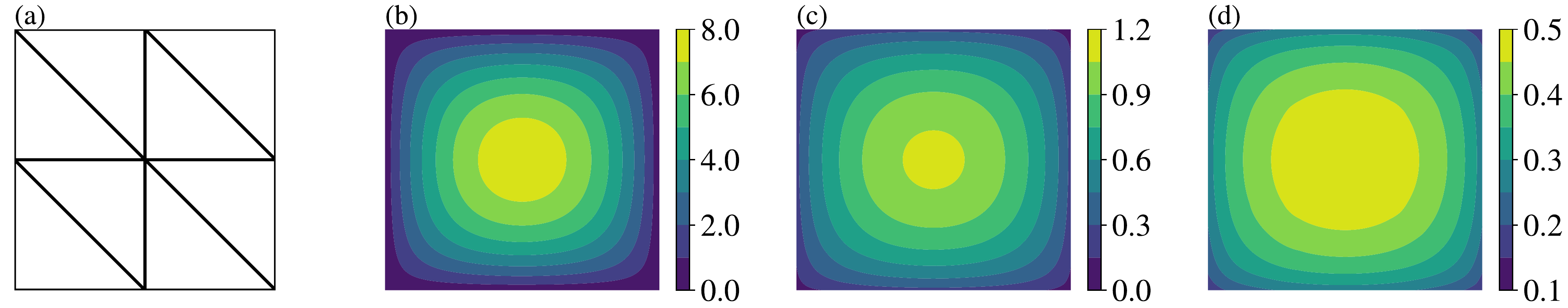}
		\par\end{centering}
	\caption{Poiseuille flows along a channel of square cross-section: (a) geometry and mesh; (b) $u_3$ contour at $\delta=100$ with $M_\text{el}=50$, $k=4$; (c) $u_3$ contour at $\delta=10$ with $M_\text{el}=50$, $k=4$; (d) $u_3$ contour at $\delta=1$ with $M_\text{el}=18$, $k=4$.}
	\label{Capert0_U}
\end{figure}

The computational performance of the HDG-SIS is investigated in the Poiseuille flow along a channel with the square cross-section of side length $H$, by comparing with solutions obtained from the same SIS but with the second-order FDM~\cite{WU2017431}. The flow is resolved on a domain of $\Omega=[0,1]\times[0,1]$. As shown in Fig.~\ref{Capert0_U}(a), the computational domain is partitioned with uniform triangles. For the discretization of velocity space, $24\times24\times24$ non-uniform points are used with a truncation of $[-4,4]$ in each direction. The typical flow velocity contours obtained by the HDG-SIS at $\delta=100$, 10 and 1 are shown in Fig.~\ref{Capert0_U}(b)-(d), respectively. It is observed that the maximum velocity emerges in the center of the flow field. As the rarefaction parameter decreases from 100 to 1, the maximum velocity reduces while the slip velocity in the vicinity of the solid surfaces increases.

For the HDG-SIS, the $L_2$ errors of the MFR, the numbers of iterative steps, and the CPU time to obtain the converged solutions are listed in Table~\ref{Carpet0_HDG}, for various numbers of triangles and degrees of approximation polynomials. The results obtained by the FDM-SIS are also listed in Table~\ref{Carpet0_FDM}, where $M_\text{p}$ denotes the number of equally-distributed discrete points in the spatial space. The $L_2$ errors are calculated using the DUGKS results as reference. For the DUKGS simulations, the same discrete velocity grid as that in the SIS is employed, while $48\times48$ and $72\times72$ points are located in the spatial space for cases with $\delta<10$ and $\delta\ge10$, respectively. Before conclusions are drawn, it should be emphasized how the CPU time is counted. At each iterative step, a global sparse linear system~\eqref{linear} needs be solved for the solution of each $\hat{h}^j$ in the HDG method. In the previous tests, the linear systems were solved by the directive solver PARDISO. Therefore, the majority of the CPU time is spent on the factorization process for the global matrices $\mathbb{K}^j$ with $j=1,\dots,M_\text{v}$. Although $\mathbb{K}^j$ varies for different discrete velocities, it does not change during the iteration. Therefore, in order to make a convictive comparison, in this test, we invert the global matrices before starting the iteration. Then, at each step, $\hat{h}^j$ are directly obtained by multiplying the inversed $\mathbb{K}^j$ to the vectors of right-hand side. Finally, $h^j$ are calculated in an element-by-element fashion. The CPU times listed here only count for the elapse of the iterations, while the time to set up the inversed global matrices is not included. Also note that storing the inversed matrices before iteration process is memory expensive, since the sparse structure is lost.


\begin{sidewaystable}
\caption{Poiseuille flow along a channel with square cross-section solved by the HDG-SIS. Itr denotes the number of iteration steps to satisfy the convergence criterion $R<10^{-5}$, and $t_\text{c}$ is the CPU time.}
\begin{centering}
\begin{tabular}{ccccccccccccccc}
\hline
\multirow{2}{*}{$k$} & \multicolumn{4}{c}{$\delta=100$} &  & \multicolumn{4}{c}{$\delta=10$} &  & \multicolumn{4}{c}{$\delta=1$}\tabularnewline
\cline{2-5} \cline{7-10} \cline{12-15}
 & $M_{\text{el}}$ & $L_{2}$ error & Itr & $t_\text{c}$, {[}s{]} &  & $M_{\text{el}}$ & $L_{2}$ error & Itr & $t_\text{c}$, {[}s{]} &  & $M_{\text{el}}$ & $L_{2}$ error & Itr & $t_\text{c}$, {[}s{]}\tabularnewline
\hline
\multirow{4}{*}{1} & 8 & $4.32\times10^{0}$ & 164 & 18.8 &  & 2 & $1.90\times10^{0}$ & 51 & 1.1 &  & 2 & $1.26\times10^{-3}$ & 14 & 0.5\tabularnewline
 & 18 & $1.83\times10^{0}$ & 128 & 63.3 &  & 8 & $4.94\times10^{-1}$ & 37 & 8.4 &  & 8 & $8.01\times10^{-3}$ & 14 & 1.7\tabularnewline
 & 32 & $9.66\times10^{-1}$ & 109 & 332.1 &  & 18 & $2.03\times10^{-1}$ & 32 & 17.8 &  & 18 & $3.40\times10^{-3}$ & 13 & 7.3\tabularnewline
 & 50 & $5.73\times10^{-1}$ & 94 & 639.6 &  & 32 & $9.85\times10^{-2}$ & 29 & 114.4 &  & 32 & $9.07\times10^{-4}$ & 13 & 24.5\tabularnewline
\hline
\multirow{4}{*}{2} & 8 & $1.41\times10^{-1}$ & 81 & 25.0 &  & 2 & $1.17\times10^{-1}$ & 30 & 2.8 &  & 2 & $6.90\times10^{-3}$ & 13 & 0.7\tabularnewline
 & 18 & $4.44\times10^{-2}$ & 69 & 153.3 &  & 8 & $2.90\times10^{-3}$ & 26 & 9.7 &  & 8 & $1.46\times10^{-3}$ & 13 & 4.1\tabularnewline
 & 32 & $1.35\times10^{-2}$ & 58 & 406.8 &  & 18 & $8.52\times10^{-3}$ & 24 & 65.6 &  & 18 & $1.71\times10^{-3}$ & 13 & 27.3\tabularnewline
 & 50 & $2.98\times10^{-3}$ & 50 & 789.5 &  & 32 & $9.29\times10^{-3}$ & 22 & 185.3 &  & 32 & $1.66\times10^{-3}$ & 13 & 61.1\tabularnewline
\hline
\multirow{4}{*}{3} & 8 & $1.78\times10^{-2}$ & 64 & 49.4 &  & 2 & $2.71\times10^{-2}$ & 28 & 4.4 &  & 2 & $3.33\times10^{-3}$ & 13 & 1.2\tabularnewline
 & 18 & $8.56\times10^{-3}$ & 49 & 230.7 &  & 8 & $1.23\times10^{-2}$ & 23 & 23.9 &  & 8 & $1.66\times10^{-3}$ & 13 & 8.5\tabularnewline
 & 32 & $6.49\times10^{-3}$ & 40 & 568.7 &  & 18 & $7.84\times10^{-3}$ & 21 & 120.7 &  & 18 & $7.84\times10^{-4}$ & 13 & 59.1\tabularnewline
 & 50 & $5.59\times10^{-3}$ & 35 & 946.5 &  & 32 & $5.74\times10^{-3}$ & 21 & 323.6 &  & 32 & $4.69\times10^{-4}$ & 13 & 123.5\tabularnewline
\hline
\multirow{4}{*}{4} & 8 & $8.49\times10^{-3}$ & 48 & 80.6 &  & 2 & $1.30\times10^{-2}$ & 23 & 3.8 &  & 2 & $1.33\times10^{-3}$ & 13 & 2.2\tabularnewline
 & 18 & $6.01\times10^{-3}$ & 37 & 287.4 &  & 8 & $7.16\times10^{-3}$ & 21 & 43.3 &  & 8 & $6.98\times10^{-4}$ & 13 & 21.0\tabularnewline
 & 32 & $5.12\times10^{-3}$ & 31 & 718.3 &  & 18 & $4.92\times10^{-3}$ & 21 & 225.5 &  & 18 & $5.72\times10^{-4}$ & 13 & 99.9\tabularnewline
 & 50 & $4.56\times10^{-3}$ & 27 & 1234.3 &  & 32 & $3.98\times10^{-3}$ & 21 & 485.9 &  & 32 & $4.92\times10^{-4}$ & 13 & 210.8\tabularnewline
\hline
\end{tabular}
\par\end{centering}
\label{Carpet0_HDG}
\end{sidewaystable}

It is found from Table~\ref{Carpet0_HDG} that for the spatial grids with the same number of triangles, the HDG-SIS solutions with higher-order of accuracy are obtained with higher degree of approximation polynomials. Therefore, to achieve the same order of accuracy, the solvers with higher degree of polynomials require spatial grids with fewer triangles. For example, when $\delta=100$, the solver with 3\textsuperscript{rd}-order polynomials has an error of about $0.8\%$ in the MFR using 18 triangles, while the one with 4\textsuperscript{th}-order polynomials reaches this accuracy with only 8 triangles. Moreover, as the rarefaction parameter decreases, fewer triangles are needed to obtained high-accuracy results. As far as the convergence speed is concerned, for all the rarefaction levels, the solvers with different degree of polynomials requires almost the same number of iterations to obtain the solutions with the same order of accuracy. For example, when $\delta=100$, about 50 steps are required to obtain MFR with $L_2$ error less than $1\%$. Since fewer triangles are needed, the higher order the solver, the less the CPU time. At $\delta=100$, the CPU time to obtain solution with $\sim0.8\%$ error with $k=4$ is about 35\% of that for the solver with $k=3$. This trend is contrary to that in an explicit DG solver, where the iterative time interval is restricted by the Counrant-Firedrichs-Lewy condition. On the same spatial grids, higher order solver requires smaller time step thus larger number of iterations to obtain converged solutions. Although the spatial grid is coarse, the large number of iterations make the 3\textsuperscript{rd}-order explicit DG kinetic solver more expensive than the 2\textsuperscript{nd}-order one for the solution of non-equilibrium flow~\cite{SU2015123}.

For the comparison of the HDG-SIS and the second-order FDM-SIS, we find that the HDG discretization is much more efficient. At $\delta=100$, the FDM predicts the MFR with error less than $1\%$ on the spatial grid with $55\times55$ discrete points, while the HDG obtained the solution with the same order of accuracy on 50, 18 and 8 triangles for $k=2$, 3 and 4 solvers, respectively. Meanwhile, at $\delta=1$, the FDM obtains the MFR with error less than $1\%$ on $35\times35$ points, while the HDG obtains the solution only on 2 triangles for all the solvers. Then, the HDG solver of $k=4$ could be $1.7$ times and more than 12 times faster than the FDM solver to obtain converged results at $\delta=100$ and $\delta=1$, respectively, according to the CPU time in Tables~\ref{Carpet0_HDG} and~\ref{Carpet0_FDM}. Although higher-order FDM could achieve higher efficiency, it demands much more efforts since stencils involving large number of points are required in the FDM scheme, which is extremely difficult to be implemented for complex geometries.

\begin{table}
\caption{Poiseuille flow along a channel of square cross-section solved by the FDM-SIS. $M_\text{p}$ is the number of discrete points in the spatial space, Itr is the number of iteration steps to satisfy the convergence criterion $R<10^{-5}$, and $t_\text{c}$ is the CPU time.}

\begin{centering}
\begin{tabular}{cccccccccccc}
\cline{1-8} \cline{10-12}
\multirow{2}{*}{$M_\text{p}$} & \multicolumn{3}{c}{$\delta=100$} &  & \multicolumn{3}{c}{$\delta=10$} &  & \multicolumn{3}{c}{$\delta=1$}\tabularnewline
\cline{2-4} \cline{6-12}
 & $L_{2}$ error & Itr & $t_{\text{{c}}}$, {[}s{]} &  & $L_{2}$ error & Itr & $t_{\text{{c}}}$, {[}s{]} &  & $L_{2}$ error & Itr & $t_{\text{{c}}}$, {[}s{]}\tabularnewline
\hline
$9^2$ & $2.22\times10^{-1}$ & 310 & 4.4 &  & $1.62\times10^{-1}$ & 51 & 1.0 &  & $7.76\times10^{-2}$ & 13 & 0.3\tabularnewline
$15^2$ & $8.66\times10^{-2}$ & 188 & 7.5 &  & $6.69\times10^{-2}$ & 38 & 1.9 &  & $3.11\times10^{-2}$ & 13 & 0.8\tabularnewline
$25^2$ & $3.49\times10^{-2}$ & 119 & 14.8 &  & $2.83\times10^{-2}$ & 30 & 4.7 &  & $1.26\times10^{-2}$ & 13 & 1.9\tabularnewline
$35^2$ & $1.99\times10^{-2}$ & 91 & 28.0 &  & $1.63\times10^{-2}$ & 29 & 11.5 &  & $7.03\times10^{-3}$ & 13 & 4.9\tabularnewline
$45^2$ & $1.36\times10^{-2}$ & 75 & 65.5 &  & $1.10\times10^{-2}$ & 32 & 32.6 &  & $4.59\times10^{-3}$ & 13 & 11.5\tabularnewline
$55^2$ & $9.36\times10^{-3}$ & 63 & 138.0 &  & $7.75\times10^{-3}$ & 26 & 79.5 &  & $2.75\times10^{-3}$ & 13 & 27.1\tabularnewline
\hline
\end{tabular}
\par\end{centering}
\label{Carpet0_FDM}
\end{table}

\subsection{Accuracy of the SIS: flows along the channels of various cross-sections}

\begin{figure}
	\begin{centering}
		\includegraphics[width=0.8\textwidth]{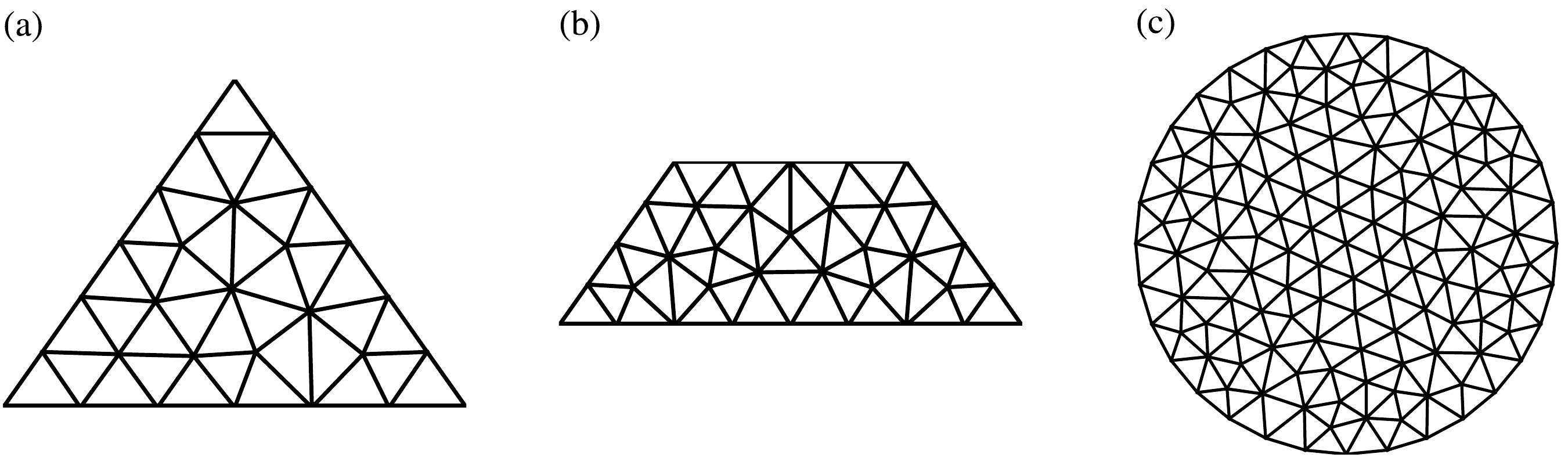}
		\par\end{centering}
	\caption{Schematics of the geometry and spatial meshes for Poiseuille flows along channels of triangular, trapezoidal, and circle cross-sections.}
	\label{Tri_G}
\end{figure}

The Poiseuille flows along the channels of triangular, trapezoidal, and circle cross-sections are used to evaluate accuracy of the HDG-SIS for flows in different geometries. Geometries and meshes are illustrated in Fig.~\ref{Tri_G}. The isosceles triangular and trapezoidal cross-sections are of acute angle $\theta=54.74^{\circ}$, and the ratio of the small and large base in the trapezoid is equal to 0.5. Totally 36, 118 and 240 triangles are used for the HDG solver with $k=3$. The molecular velocity space is discretized in the range of $[-4,4]$ by 32 non-uniform points in each direction. The characteristic length $H$ for the flow in triangular and trapezoidal cases is set as its hydraulic diameter, i.e. 4 times the ratio of area and perimeter. In the circle case, the radius is chosen as the characteristic length.

\begin{figure}
	\begin{centering}
		\includegraphics[width=0.95\textwidth]{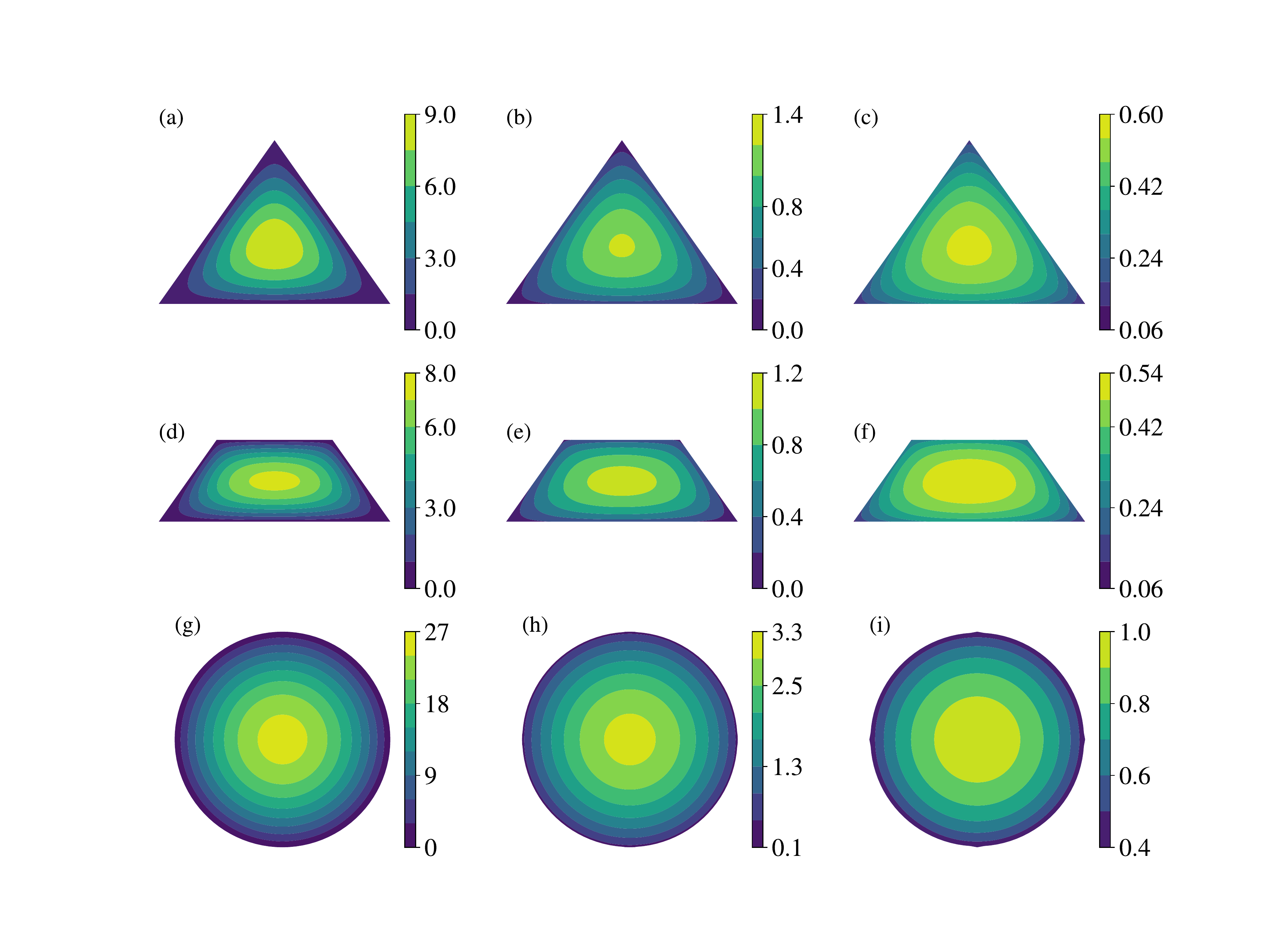}
		\par\end{centering}
	\caption{Velocity contours in the Poiseuille flows along the channels of triangle, trapezoid, and circle cross-sections. The rarefaction parameters in the first, second, and third columns are  $\delta=100$, 10, and 1, respectively. }
	\label{Tri_U}
\end{figure}

\begin{figure}
\begin{centering}
\includegraphics[width=0.55\textwidth]{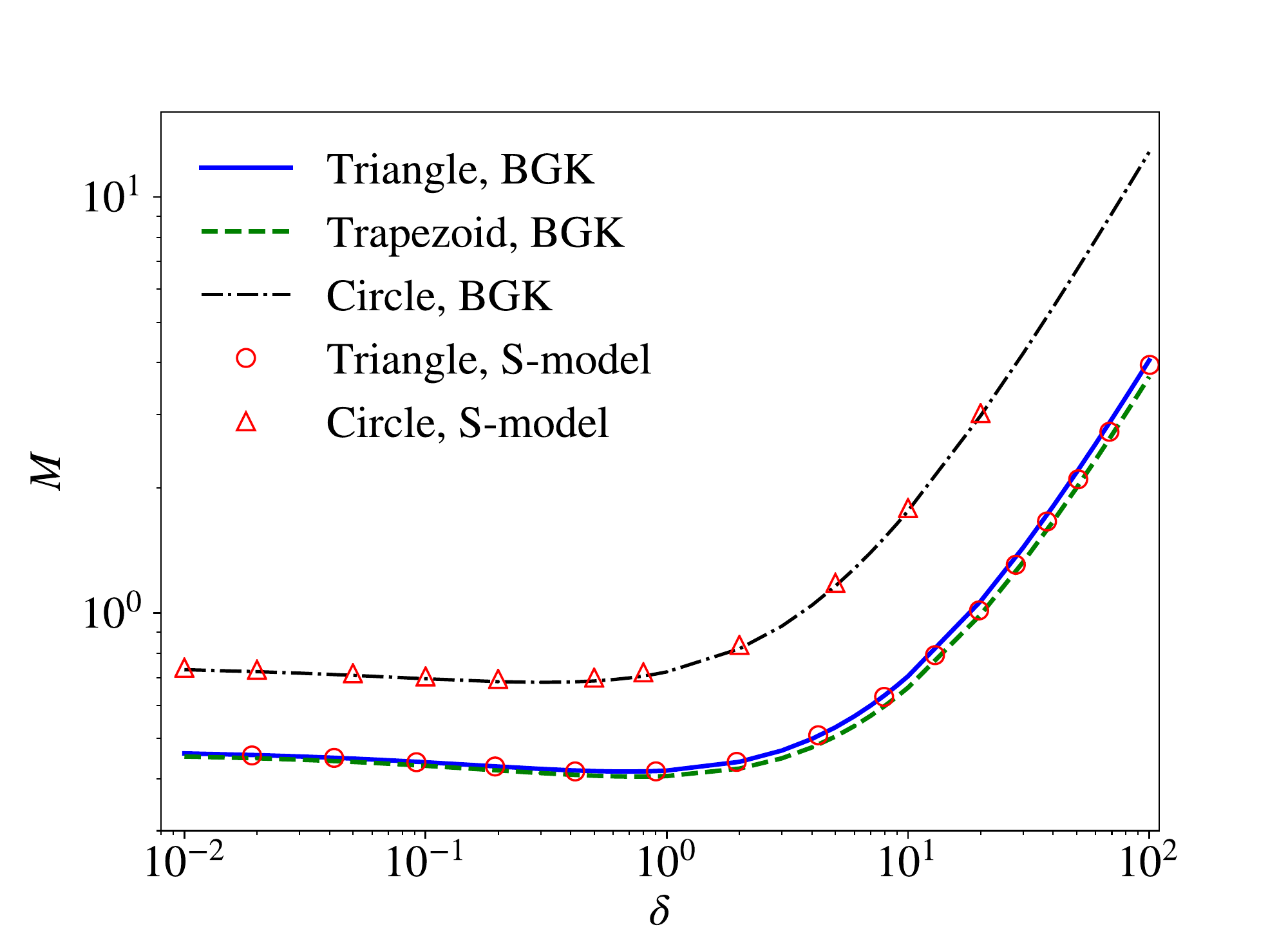}
\par\end{centering}
\caption{MFRs of the Poiseuille flow along the channels of triangle, trapezoid, and circle cross-sections. the lines are solutions from the HDG-SIS solver based on the linearized BGK kinetic model, the symbols are solutions from the FDM solver based on the linearized Shakhov kinetic model (S-model). }
\label{Tri_M}
\end{figure}

Velocity contours at $\delta=100$, $10$, and $1$ are shown in Fig.~\ref{Tri_U}. Similar to flows in the square channel, the maximum velocities appear in the center of the flow field, which decrease as the rarefaction parameter decreases. MFRs over a wide range of degree of rarefaction are plotted in Fig.~\ref{Tri_M} and compared to the data from Ref.~\cite{Graur2009,Ritos2011}. The MFRs for the triangular and trapezoidal channels are close to each other due to the fact that the hydraulic diameter is chosen as the characteristic length to nondimensionalize the problem. If using the radius as the characteristic length, the MFR in circle channel is larger than those in the other two channels. The Knudsen minimum, where the MFR is minimum, also arises at $\delta\sim1$. In all cases, the HDG-SIS results agree well with those in literature, which demonstrates the accuracy of the proposed HDG-SIS scheme. It is worth to mention that the results in literature were calculated from the linearized Shakhov kinetic model equation, where the additional correction of the heat flux in the collision operator, actually has no effect on the MFR.

\subsection{Capability to handle complex geometry: flows along Apollonian gasket channels}

\begin{figure}[t]
	\begin{centering}
		\includegraphics[width=0.9\textwidth]{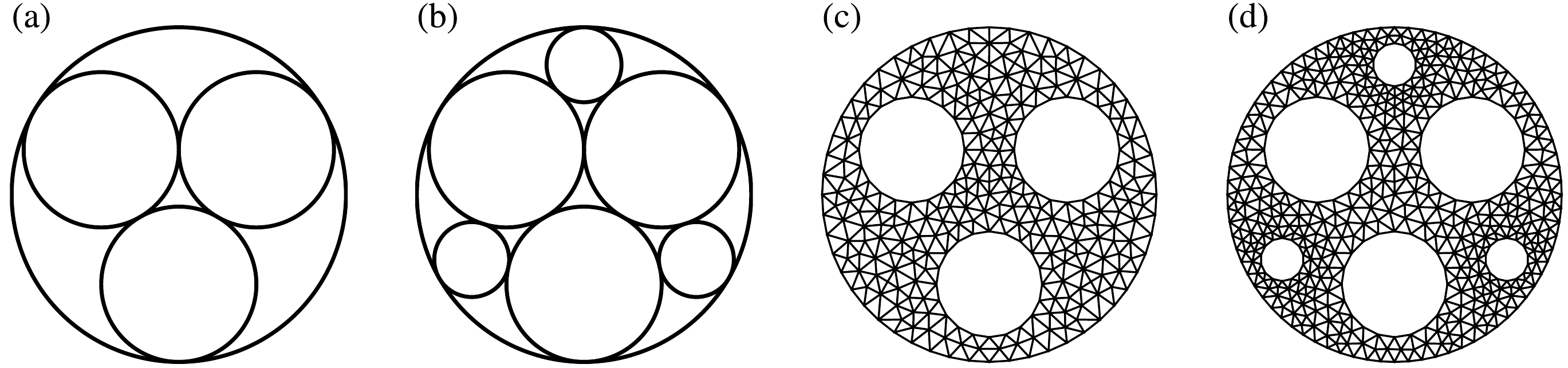}
		\par\end{centering}
	\caption{(a)-(b): Schematics of the geometry for the original Level-1 and Level-2 Apollonian gaskets. (c)-(d): Schematics of the geometry and mesh for the Poiseuille flows along the Level-1 and Level-2 Apollonian gasket channels.}
	\label{Apo_G}
\end{figure}

Finally, the Poiseuille flow along the channels with cross-section described by the Apollonian fractal gasket is used to demonstrate capability of the proposed HDG-SIS to handle complex geometries. The cross-section of the original Apollonian fractal gasket is a fractal generated starting from a circle, which is filled in a triple of circles with the same radius, each is tangent to the other three (including the internal tangent with the outer circle, see Fig.~\ref{Apo_G}(a)). Then for the next level, the structure is filled in 3 more circles, each is tangent to another three (see Fig.~\ref{Apo_G}(b)). Here, for the geometry we calculated, the inner circles are not tangent to anyone of the others, while their centers coincide to those in the original Apollonian gaskets and their radii are determined such that the porosity (the fraction of the area of voids over the total area) is 0.7 for the first level and on this basis, the porosity of the second level is 0.65. The resulting geometries and meshes of the Level-1 and Level-2 structures are presented in Fig.~\ref{Apo_G}(c)-(d). In the current simulation, we treat the inner small circles as solids and the flow flowing through the gaps between the outer circle and the inner ones. To determine the rarefaction parameter, the radius of the outer circle is set as the characteristic length for non-dimensionlization. Totally 494 and 1082 triangles are employed in spatial discretization and the velocity grid is the same as the previous tests.

\begin{figure}
	\begin{centering}
		\includegraphics[width=0.9\textwidth]{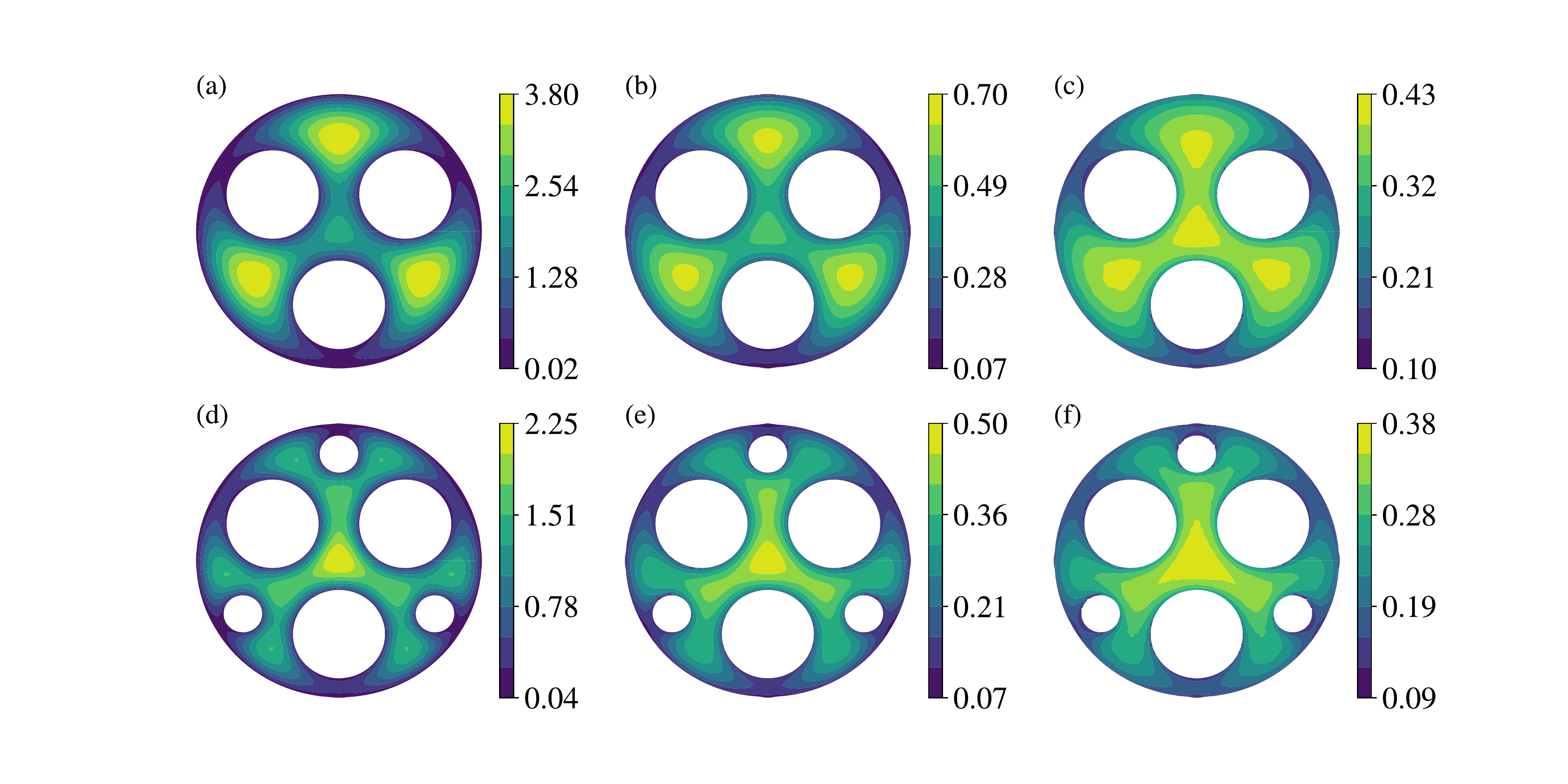}
		\par\end{centering}
	\caption{Velocity contours in the Poiseuille flows along the Apollonian gasket channels. The rarefaction parameters in the first, second, and third columns are  $\delta=100$, 10, and 1, respectively.}
	\label{Apo_U}
\end{figure}

\begin{figure}
\begin{centering}
\includegraphics[width=0.55\textwidth]{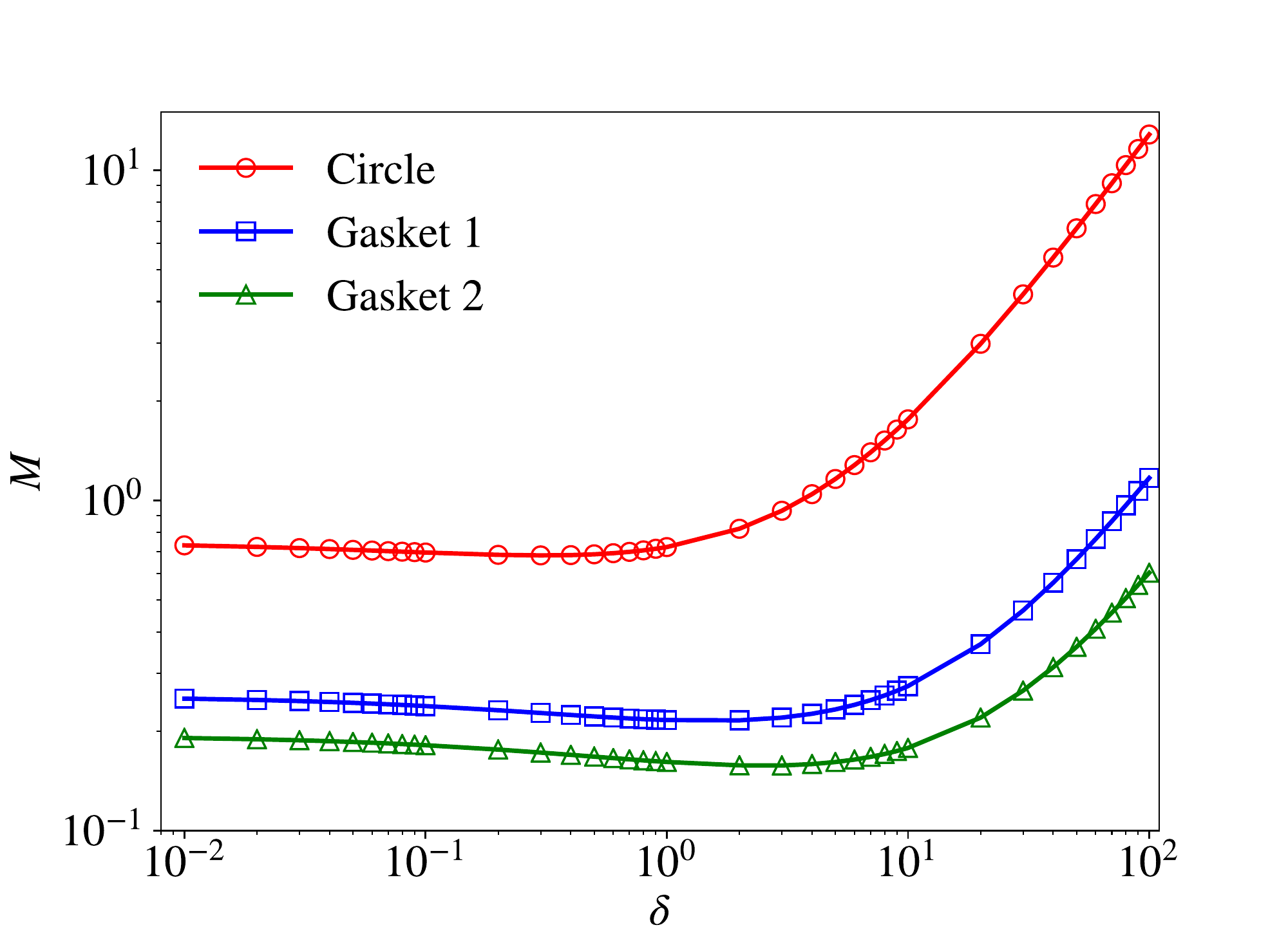}
\par\end{centering}
\caption{MFRs of the Poiseuille flow through the Apollonian gasket channels solved by the HDG-SIS.}
\label{Apo_M}
\end{figure}

Figure~\ref{Apo_U} displays the velocity contours in the different geometries with varying rarefaction parameters, where the velocity distributions possess an axial symmetry. When there is no solid inside the outer circle (the last row in Fig.~\ref{Tri_U}), the maximum velocity is at the center of the domain. However, for the Level-1 geometry, the large flow velocities move along the radial direction to the outer boundary. While for the Level-2 geometry, the large flow velocities emerge in the center of the domain again. Fig.~\ref{Apo_M} shows the MFRs in the Poiseuille flow through the Apollonian gasket channels together with the one through the circle channel. As the recursion level increases, the porosity of the Apollonian gasket channel decreases, so as the MFR. The Knudsen  minimum in the MFR can be seen, however, the location of the minimum MFR shift towards larger values of $\delta$ in the Apollonian gasket channels compared to the one in the circle channel. This is because, in the calculation of $\delta$ the characteristic flow length $H$ is selected to be the radius of the outer circle, which is larger than the radius of the solid near which the flow velocity is maximum.

\section{Conclusions}\label{Concludsion}

In summary, based on the high-order hybridizable discontinuous Galerkin discretization, we have developed an accurate and efficient numerical method to find the steady-state solution of the linearized BGK model equation, for rarefied Poiseuille gas flow through the channels with cross-sections of arbitrary shape. First, an HDG solver with approximation polynomial of degree up to 4 has been developed. The discrete perturbed molecular velocity distribution functions and their traces are approximated on arbitrary triangular mesh and the mesh skeleton, respectively. Based on the first-order upwind scheme, a numerical flux has been designed to evaluate the convection between adjacent cells. By imposing the continuity of the normal flux, a final global systems for VDF traces are obtained. Since the traces are defined on the cell interfaces and have single-values, the global coupled degrees of freedom of the unknowns are significantly reduced compared to the classical DG method. The boundary condition has been implemented equivalently to the standard Neumann boundary condition. In this way, the boundary condition could be treated in a unified framework the same as the calculation of flux on interfaces.

In parallel to the HDG solver for the gas kinetic equation, a macroscopic diffusion equation for flow velocity is synchronously solved on the same mesh. At each iterative step, the VDF in the bulk region is corrected by the flow velocity from the diffusion equation. Since the macroscopic equation boots the exchange of information, fast convergence with asymptotic-preserving into the fluid dynamic limit is realized for the steady-state solution within the near-continuum flow regime. On the other hand, high-order moments of VDF in the diffusion equation preserve the accuracy of the scheme in highly rarefied gas flows.

Four different validation problems of the Poiseuille flow along long channels with various cross-sections have been presented to show accuracy and capability of the proposed scheme. Several conclusions have been obtained through the performance analysis:
\begin{itemize}
\item{Compared to the conventional iterative scheme, the synthetic iterative scheme can significantly reduce the number of iterative steps to reach the steady-state solution in near-continuum flow regimes: the synthetic iterative scheme could be more than 100 times faster.}

\item{To obtained the results with the same order of accuracy, the HDG solver with higher degree of approximation polynomial requires fewer triangles in spatial mesh. As a result, the computational time and memory consumption can be further reduced.}

\item{Compared to the synthetic iterative scheme solved by FDM, the HDG discretization is much more efficient. To obtained the results with the same order of accuracy, the HDG scheme can be faster than the FDM by one order of magnitude.}
\end{itemize}

It is worth mentioning that the basic hybridizable DG formulation developed in this paper is not limited to the linearized BGK equation. It is straightforward to be extended for other gas kinetic model equations, or even the full Boltzmann equation by adopting a proper method (e.g. fast spectral method~\cite{WU2013} and conservative projection method~\cite{TCHEREMISSINE1998}) to calculate the Boltzmann collision operator. Since the computational cost of the Boltzmann collision operator is much higher than that of the gas kinetic models, and the HDG with higher degree of approximation polynomial can reduce the spatial triangular meshes (and hence the nodal points where the Boltzmann collision operator is evaluated), the advantage of using the HDG method will become more obvious. Also, the HDG-SIS is ready to be extended for rarefied gas mixtures.

\section*{Acknowledgments}

This work is jointly founded by the Royal Society of Edinburgh and National Natural Science Foundation of China under Grant No. 51711530130. It is also financially supported by the Carnegie Research Incentive Grant for the Universities in Scotland, and the Engineering and Physical Sciences Research Council (EPSRC) in the UK under grant EP/M021475/1.

\appendix

\section*{Appendix}
\renewcommand{\theequation}{A.\arabic{equation}}

Here, some details for implementing the HDG method for the linearized BGK model equation are presented. The weak form of the HDG local and global problems are:
\begin{equation}
-\left(\nabla\varphi,\bm v^jh^j\right)_{\Omega_i}+\sum^3_{e=1}\langle\varphi,\left(\bm v^j\cdot\bm n-\alpha\right)\hat{h}^j\rangle_{\partial\Omega^e_i}+\sum^3_{e=1}\langle\varphi,\alpha h^j\rangle_{\partial\Omega^e_i}+(\varphi,\delta h^j)_{\Omega_i}=(\varphi,s^j)_{\Omega_i},
\label{LPA}
\end{equation}
for $i=1,\dots,M_\text{el},\ j=1,\dots,M_\text{v}$, and
\begin{equation}
\begin{aligned}
\langle\psi,\hat{h}^j\rangle_{\Gamma_c}=\frac{1}{2}\langle\psi,h^j_{\eta(c,+)}+h^j_{\eta(c,-)}\rangle_{\Gamma_c},\quad\text{on}\ \Gamma\backslash\partial\Omega,\\
\langle\psi,\hat{h}^j\rangle_{\Gamma_c}=\frac{1}{2}\langle\psi,h^j_{\eta(c,+)}+g^j\rangle_{\Gamma_c},\quad\text{on}\ \Gamma\cap\partial\Omega,
\end{aligned}
\label{GPA}
\end{equation}
for $j=1,\dots,N$. The local problem (\ref{LPA}) represents a system of equations for each triangle $\Omega_i$ and discrete velocity $v^j$, which allows unknown $h^j$ as a function of the trace unknown $\hat{h}^j$. Then, replaced in Eq. (\ref{GPA}), a global system is set up in terms of only the unknown trace.

In this paper, unknowns are approximated by nodal shape functions $N_l$ in each triangle $\Omega_i$ or by $\hat{N}_l$ on each face $\Gamma_c$, which have the form given below:
\begin{equation}
\begin{aligned}
h^j_i=\sum^{K_\text{el}}_{l=1}N^l_iH^j_{i,l},\quad\text{in}\ \Omega_i\\
\hat{h}^j_c=\sum^{K_\text{fc}}_{l=1}\hat{N}^l_c\hat{H}^j_{c,l},\quad\text{on}\ \Gamma_c
\end{aligned}
\end{equation}
where $K_\text{el}={(k+1)(k+2)}/{2}$ and $K_\text{fc}=k+1$ are the numbers of degree of freedom, when the approximations are sought in the finite element space of polynomials of degree up to $k$. If we denote the $\mathbf{H}^{i,j}$ as the vector of nodal value of $h^j$ on each triangle $\Omega_i$, $\mathbf{\hat{H}}^{i,j}$ as the vector summing all the nodal value of $\hat{h}^j$ on the 3 faces of triangle $\Omega_i$, and $\mathbf{\hat{H}}^{c,j}$ as the vector of nodal value of $\hat{h}^j$ on each face $\Gamma_c$, both the local and global problem can be rewritten in the matrix form as:
\begin{equation}
\mathbf{H}^{i,j}=\left[\mathbf{A}^{i,j}\right]^{-1}\mathbf{S}^{i,j}+\left[\mathbf{A}^{i,j}\right]^{-1}\mathbf{\hat{A}}^{i,j}\mathbf{\hat{H}}^{i,j},
\label{LPAA}
\end{equation}
and
\begin{equation}
\begin{aligned}
\mathbf{\hat{B}}^{c,j}\mathbf{\hat{H}}^{c,j}=\mathbf{B}^{\eta(c,+),j}\mathbf{H}^{\eta(c,+),j}+\mathbf{B}^{\eta(c,-),j}\mathbf{H}^{\eta(c,-),j},\quad\text{on}\ \Gamma\backslash\partial\Omega,\\
\mathbf{\hat{B}}^{c,j}\mathbf{\hat{H}}^{c,j}=\mathbf{B}^{\eta(c,+),j}\mathbf{H}^{\eta(c,+),j}+\mathbf{\hat{S}}^{c,j},\quad\text{on}\ \Gamma\cap\partial\Omega,
\end{aligned}
\end{equation}
where
\begin{equation}
\begin{aligned}
\mathbf{A}^{i,j}_{ml}=\delta\left(N^m_i,N^l_i\right)_{\Omega_i}+\sum^3_{e=1}\alpha\langle N^m_i,N^l_i\rangle_{\partial\Omega_i}-\left(\bm v^j\cdot\nabla N^m_i,N^l_i\right)_{\Omega_i},\\
\mathbf{\hat{A}}^{i,j,e}_{ml}=\left(\alpha-\bm v^j\cdot\bm n\right)\langle N^m_i,\hat{N}^l_{\sigma(i,e)}\rangle_{\partial\Omega^e_i},\\
\mathbf{S}^{i,j}_m=\left(N^m_i,s^j\right)_{\Omega_i},\\
\mathbf{\hat{B}}^{c,j}_{ml}=\langle\hat{N}^m_c,\hat{N}^l_c\rangle_{\Gamma_c},\\
\mathbf{B}^{\eta(c,\pm),j}_{ml}=\frac{1}{2}\langle\hat{N}^m_c,N^l_{\eta(c,\pm)}\rangle_{\Gamma_c},\\
\mathbf{\hat{S}}^{c,j}_{m}=\frac{1}{2}\langle\hat{N}^m_c,g^j\rangle_{\Gamma_c}.\\
\end{aligned}
\end{equation}
By eliminating the unknowns $\mathbf{H}^{i,j}$ with Eq.~\eqref{LPAA} and assembling the equations of global problem over all the faces, the global problem becomes
\begin{equation}
\mathbb{K}^j\mathbf{H}^j=\mathbb{R}^j,
\label{GPAA}
\end{equation}
where $\mathbf{\hat{H}}^j$ is the vector of nodal value of $\hat{h}^j$, which is sum of all the faces in the computational domain, and,
\begin{equation}
\begin{aligned}
\mathbb{K}^j=\A^{M_\text{fc}}_{c=1}\mathbf{\hat{B}}^{c,j}-\mathbf{B}^{\eta(c,\pm),j}\left[\mathbf{A}^{\eta(c,\pm),j}\right]^{-1}\mathbf{\hat{A}}^{\eta(c,\pm),j},\\
\mathbb{R}^j=\A^{M_\text{fc}}_{c=1}\mathbf{B}^{\eta(c,\pm),j}\left[\mathbf{A}^{\eta(c,\pm),j}\right]^{-1}\mathbf{S}^{\eta(c,\pm),j}+\mathbf{\hat{S}}^{c,j}.
\end{aligned}
\end{equation}
To obtained the global matrix $\mathbb{K}^j$ and vector $\mathbb{R}^j$, the dense matrices $\mathbf{A}^{i,j}$ with dimension $K_\text{el}\times K_\text{el}$ for each $i=1,\dots,M_\text{el},\ j=1,\dots,M_\text{v}$ need to invert. Then, the sparse unsymmetric linear systems of equations (\ref{GPAA}) is directly solved to determine $\mathbf{\hat{H}}^j$. Finally, $\mathbf{H}^{i,j}$ is updated in an element-by-element fashion respecting to Eq. (\ref{LPAA}).

\section*{References}

\bibliographystyle{elsarticle-num}
\bibliography{mybibfile}

\end{document}